\documentclass[twocolumn,longauthor]{aastex631}
\pdfoutput=1

\newcommand{\athena}{\textit{Athena\texttt{++}}}
\newcommand{\bs}{\boldsymbol}
\newcommand{\lowbeta}{{\tt\string b5}}
\newcommand{\highbeta}{{\tt\string b200}}
\newcommand{\lowbetasup}{{\tt\string b5\_hi}}
\newcommand{\highbetasup}{{\tt\string b200\_hi}}
\newcommand{\llangle}{\ensuremath{\langle\langle}}
\newcommand{\rrangle}{\ensuremath{\rangle\rangle}}
\newcommand{\Epsilon}{\mathcal{E}}

\usepackage{amsmath}
\usepackage[T1]{fontenc}

\accepted{October 13, 2023}

\submitjournal{The Astrophysical Journal}

\shorttitle{Magnetic field evolution in high and low $\beta$ disks with initially-toroidal fields}
\shortauthors{Rodman \& Reynolds}

\begin{document}

\title{Magnetic field evolution in high and low $\beta$ disks with initially-toroidal fields}

\correspondingauthor{Payton E. Rodman}
\email{per29@cam.ac.uk}

\author[0000-0002-1624-9359]{Payton E. Rodman}
\affiliation{Institute of Astronomy, University of Cambridge, Madingley Rd, Cambridge CB3 0HA, UK}

\author[0000-0002-1510-4860]{Christopher S. Reynolds}
\affiliation{Institute of Astronomy, University of Cambridge, Madingley Rd, Cambridge CB3 0HA, UK}
\affiliation{Dept. of Astronomy \& Joint Space Science Institute (JSI), University of Maryland, College Park, MD~20742}

\begin{abstract}
We present results from a pair of high resolution, long timescale ($\sim10^5~GM/c^3$), global, three dimensional magnetohydrodynamical accretion disk simulations with differing initial magnetic plasma $\beta$ in order to study the effects of initial toroidal field strength on production of large-scale poloidal field. We initialize our disks in approximate equilibrium with purely toroidal magnetic fields of strength $\beta_0=5$ and $\beta_0=200$. We also perform a limited resolution study. We find that simulations of differing field strength diverge early in their evolution and remain distinct over the time studied, indicating that initial magnetic conditions leave a persistent imprint in our simulations. Neither simulation enters the Magnetically Arrested Disk (MAD) regime. Both simulations are able to produce poloidal fields from initially-toroidal fields, with the $\beta_0=5$ simulation evolving clear signs of a large-scale poloidal field. We make a cautionary note that computational artifacts in the form of large-scale vortices may be introduced in the combination of initially-weak field and disk-internal mesh refinement boundaries, as evidenced by the production of an $m=1$ mode overdensity in the weak field simulation. Our results demonstrate that the initial toroidal field strength plays a vital role in simulated disk evolution for the models studied.
\end{abstract}


\section{Introduction} \label{sec:intro}
Accretion disks are highly turbulent and complex astrophysical systems formed by infalling gas circularizing around a massive central body such as a young star, a white dwarf, a neutron star, or a black hole. In the case of accretion onto a supermassive black hole, the disk physics governs both the growth of the black hole and the form of the resulting energy output and thus has wide-reaching implications not just for the immediate environment but on galaxy scales \citep[e.g. the $M$-$\sigma$ relation;][]{King2003}. The strong gravitational forces, relativistic velocities, and magnetic fields present make these accreting black hole systems an invaluable laboratory for studying a wide range of extreme physical processes, including the properties of strong gravitational fields, particle acceleration, and the interaction of magnetic fields and fluid turbulence, the last of which forms the primary focus of this paper.

For an accretion disk to actually accrete it must find a way to remove angular momentum from orbiting fluid elements. This could in theory occur through local exchanges between gas parcels within the disk and/or through a wholesale loss of angular momentum from the system via, for example, magnetic braking from a disk wind. Given the extremely high Reynolds numbers of most astrophysical disks, intrinsic microscopic viscosity can be immediately ruled out as an angular momentum transport mechanism as it would drive accretion over timescales that are far too long to explain the observed luminosity of disks \citep[see][for derivation]{Balbus2003}. The seminal work of \citet{ShakuraSunyaev1973} acknowledged that a form of anomalous viscosity could exist as a result of turbulence, potentially related to the presence of magnetic fields, but the mechanism for generating this turbulence was then unable to be specified. Simply invoking a very high Reynolds number as a driver of turbulence is not sufficient as Keplerian disks are linearly and non-linearly stable to shearing instabilities even at very high Reynolds numbers \citep{HawleyEA1995,HawleyEA1996,BalbusEA1996}. Later work by \citet{BalbusHawley1991,BalbusHawley1998} identified a powerful magnetohydrodynamic (MHD) instability which can drive the necessary turbulence. This magneto-rotational instability (MRI) has since become the paradigm for accretion disk theory.

Magnetohydrodynamic (MHD) and (more recently) General Relativistic Magnetohydrodynamic (GRMHD) simulations are now common tools for studying the internal dynamics of the disk due to the highly non-linear nature of the MRI-driven turbulence \citep[see e.g.][for review]{Balbus2003,DavisTchekhovskoy2020}. In spite of much attention in recent years, however, many questions still remain about the evolution of magnetic fields within accretion disks and the nature of the turbulence that is formed. Equally importantly, uncertainties remain about the influence of computational realities (finite spatial resolution, memory of initial conditions in a limited temporal span, and non-helicity preserving MHD algorithms) on the outcome of accretion disk simulations.

Choices in the initial vertical magnetic field were shown to have important implications for turbulence and overall disk evolution even in early simulations \citep{HawleyEA1995}, with turbulent stresses increasing with increasing net vertical field. Later studies of both shearing box \citep[e.g.][]{SalvesenEA2016,BaiStone2013} and global \citep[e.g.][]{MishraEA2020} simulations showed that the presence of an initially vertical field affects the structure of the simulated accretion flow. If the vertical field in the inner regions of the disk is sufficiently strong, it may cause the disk to enter a Magnetically Arrested Disk \citep[MAD;][]{NarayanEA2003} state in which mass accretion is suppressed or even temporarily halted due to magnetic pressure and, if the central black hole is spinning, jets are launched via the Blandford-Znajek mechanism \citep{BlandfordZnajek1977,TchekhovskoyEA2011}. MADs have been invoked to explain the low-luminosity and powerful jet in M87 and GRMHD models of MADs can quantitatively match the horizon-scale mm-band image obtained by the Event Horizon Telescope \citep{EHTCollab2019}. The counterpart to MAD is the Standard And Normal Evolution (SANE) state, which is effectively characterized by the absence of a MAD state. Recent work by \citet{BegelmanEA2022}, however, suggests that while the presence of a {\it net} vertical field is important, it plays a secondary role to that of the toroidal field, with the toroidal component being critically important in the saturation of magnetic flux. To add further complexity to this discussion, the net vertical magnetic flux may be zero on a global scale but {\it non-zero} locally \citep{SorathiaEA2010}. A zero-net-vertical-flux (ZNVF) global disk may well be composed of many net-vertical-flux (NVF) local ``shearing box'' patches, magnetically connected at high-altitudes through the low density corona of the disk.

In addition to questions on the role of vertical fields and fluxes, the final state of a simulation appears to depend significantly on the particular initial field morphology used, even where the initial net vertical field is kept constant. Early work by \citet{MachidaEA2001} suggested that the initial magnetic field configuration could affect the mass outflow rate in Radiatively-Inefficient Accretion Flows (RIAFs), disks in which the radiative cooling is negligible and local cooling driven by advection --- such disks are generally geometrically-thick and have low-accretion rate. Such flows  lack a large separation of scales between their inflow/viscous times, thermal times, and dynamical times and so seem to retain memory of their initial conditions during the inflow \citep{WhiteEA2020}. Further, since they lack radiative cooling, the sign of the Bernoulli parameter (and hence the degree to which regions of the disk are unbound) is sensitive to both the feeding prescription and the detailed internal energetics associated with the initial field. Later simulations by \citet{NarayanEA2012} found that a MAD state could be produced by varying the number and direction of poloidal loops in the initial magnetic field, with a single-loop configuration being most conducive to building up vertical field necessary for the MAD state. \citet{WhiteEA2020} then built upon this work, showing that within SANE models variations in initial magnetic field configuration have strong effects on the energetics and structure of RIAFs, with smaller initial poloidal loops generating a less coherent radial field (leading to less horizon-penetrating magnetic flux) and producing polar inflows while larger poloidal loops showed the opposite. Similar work by \citet{JiangEA2019} on thin, radiation-dominated disks also found factors of three differences in accretion rate and vertical disk structure depending on whether the initial poloidal field was dipolar or quadrupolar , while \citet{MishraEA2022} found that thin RIAFs initialized with dipolar poloidal fields were thermally unstable and those with quadrupolar oscillated between stable and unstable states.

These combined results clearly demonstrate that the properties of simulated accretion disks are dependent on the artificial and somewhat arbitrary initial magnetic conditions used in their construction, and such a dependence raises serious questions as to the validity of simulations as emulations of real astrophysical disks.

Magnetic fields with net vertical flux or with significant large-scale vertical/polar components play an important role not only for the emergence of MAD states, but also for the production of astrophysical jets through the Blandford-Znajek \citep{BlandfordZnajek1977} and Blandford-Payne \citep{BlandfordPayne1982} processes. An open question remains whether, and under what conditions, a disk can eventually forget the details of the initial field configuration and grow a self-consistent large-scale field via dynamo action given sufficient time. \citet{FragileSadowski2017} found that initially strongly magnetized disks (where the ratio of thermal-to-magnetic pressure is only $\beta=P_{\rm gas}/P_{\rm mag}\sim10$) with purely toroidal fields were not locally self-sustaining, albeit the simulations were run over a short timescale ($700~GM/c^3$) that may not have allowed enough time for poloidal growth. Global simulations by \citet{LiskaEA2020} run for much longer times ($1.3\times10^5~GM/c^3$) successfully generated a strong enough poloidal field {\it in situ} to enter the MAD regime, but only from an initially strong field ($\beta=5$), which is likely not reflective of initial conditions in nature.

Given the long times needed for a disk with an initially weak toroidal magnetic field to reach equilibrium out to reasonable radii, most simulations with initially toroidal fields are initialized with a low magnetic plasma $\beta$ (i.e. a `strong' magnetic field), meaning little is known of the amplification process for $\beta>100$ and few simulations cover the intersection of weak and toroidal initial field configurations. 

In this work, we study the ability of simulated disks initialized with a purely toroidal field to grow a large-scale poloidal field. We agree with and extend the previous findings of \citet{LiskaEA2020}, finding large-scale field growth for disks initialized with either a strong ($\beta=5$) or a weak ($\beta=200$) magnetic field. The core of this work is based on two long-duration simulations with initially-toroidal fields, a strong-field ($\beta=5$) and a weak-field ($\beta=200$) case. Our longest simulation (the weak-field case) is run out to $t\sim4\times10^5~GM/c^3$ in order to capture the long-term evolution of the magnetic field. Given our focus on understanding the influence of numerics on results, we also run two additional simulations that have identical set-up except for higher resolution. We see evidence of poloidal flux generation in all cases, with the weaker initial field case saturating with weaker and more disorganized poloidal field. All of our model disks display dynamo-cycles in the toroidal field even though the disks are geometrically-thick. We also identify the subtle ability of internal mesh-refinement boundaries to drive vorticity-induced density asymmetries. 

The basic equations of the problem and computational setup are given in Section~\ref{sec:model}. The analysis of our four simulations, with special emphasis on the growth of large-scale poloidal field, are presented in Section~\ref{sec:results}. We discuss and summarize our findings with our conclusions in Section~\ref{sec:discussion}.

\section{Basic Model and Computational Setup} \label{sec:model}

In this paper, we model the evolution of two global accretion disks with initially toroidal magnetic fields, one weak-field case and one strong-field case (see Section~\ref{subsec:sim_setup}). We additionally produce two more simulations with the same initial magnetic field strengths but with double resolution in the $\theta$- and $\phi$-directions. As the goal of this work is to study the evolution of the magnetic field, we keep all other aspects fixed between simulations. To reduce the computational resources needed, we neglect any physics which does not contribute directly to the evolution of the magnetic field; specifically this work does not include radiative physics (formally making our disks RIAFs) nor full general relativity (instead employing a pseudo-Newtonian potential). Instead, we focus all resources into evolving the equations of non-relativistic ideal-MHD at high resolution for as long as possible.

All lengths, times, and velocities are given in natural units scaled to the mass of the black hole, formally setting $G=M=c=1$.

\subsection{Governing Equations and Simulation Code}\label{subsec:sim_code}
Our accretion disk models are evolved using the MHD code \athena{}~\textit{v21.0} \citep{athena++} on a spherical polar grid $(r,\theta,\phi)$, with the piece-wise linear method (PLM) for spatial reconstruction, second-order (VL2) time-stepping, a HLLD Riemann solver, and a Courant, Friedrichs, \& Lewy (CFL) number of $0.3$. The code is parallelized over 1008 Cascade Lake cores on the Cambridge Service for Data Driven Discovery (CSD3) system {\itshape Peta4}, achieving a full-node performance of $\sim$~21-22 MZone-cycles/core-second. 

\athena{} solves the equations of ideal MHD, given in conservative form as \citep{athena}
\begin{align}\label{eqn:mhd}
    \frac{\partial\rho}{\partial t} + \nabla\cdot\left(\rho\bs{v}\right) &= 0,\\
    \frac{\partial\rho\bs{v}}{\partial t} + \nabla\cdot\left(\rho\bs{vv} - \bs{BB} + {\bs P^*}\right) &= -\rho\nabla\Phi_{\rm PW},\\
    \frac{\partial E}{\partial t} + \nabla\cdot\left[\left(E + P^*\right){\bs v} - {\bs B}\left(\bs{B\cdot v}\right)\right] &= 0,\\
    \frac{\partial{\bs B}}{\partial t} - \nabla\times\left({\bs v}\times{\bs B}\right) &= 0,
\end{align}
where $\bs{P^*}=(P + B^2/2)\bs{I}$, $\bs{I}$ is the diagonal unit tensor, $B^2 = \bs{B}\cdot\bs{B}$, and all other variables having their usual meaning. The total energy density is given by
\begin{equation}
    E = \frac{P}{\gamma - 1} + \frac{1}{2}\rho v^2 + \frac{B^2}{2}+\rho\Phi_{\rm PW},
\end{equation}
where we have imposed a $\gamma$-law equation of state relating the internal energy density $\Epsilon$ to the pressure, $P=(\gamma-1)\Epsilon$. We use the pseudo-Newtonian Paczy\'{n}ski–Wiita gravitational potential, $\Phi_{\rm PW}$,
\begin{equation}
    \Phi_{\rm PW}(r) = - \frac{GM}{r-2r_g}, \qquad r_g \equiv GM/c^2,
\end{equation}
which features an innermost stable circular orbit (ISCO) at $r_{\rm ISCO} = 6r_g$ and is an adequate approximation (within Newtonian physics) of the gravitational field around a non-rotating black hole. The solenoidal condition on the magnetic field, 
\begin{equation}\label{eqn:solenoidal}
\nabla\cdot\bs{B}=0,
\end{equation} 
is enforced by upwind constrained transport \citep[CT;][]{EvansHawley1988}.

The system of equations \ref{eqn:mhd}--\ref{eqn:solenoidal} is closed and may be solved numerically. During the numerical solution, we enforce density and pressure floors at $\rho_{\rm floor} = 3\times10^{-5}$ and $P_{\rm floor} = 1\times10^{-10}$. Due to the high Alfv\'{e}n speeds that arise in the low-density regions of our \lowbetasup\ run, we increased the density floor in this simulation to $\rho_{\rm floor} = 3\times10^{-4}$ at $t\sim2\times10^4~GM/c^3$, representing a $0.16$\% increase in the total mass at that moment which does not appreciably affect our results.

\subsection{Simulation Setup}\label{subsec:sim_setup}

\begin{figure}[t!]
    \plotone{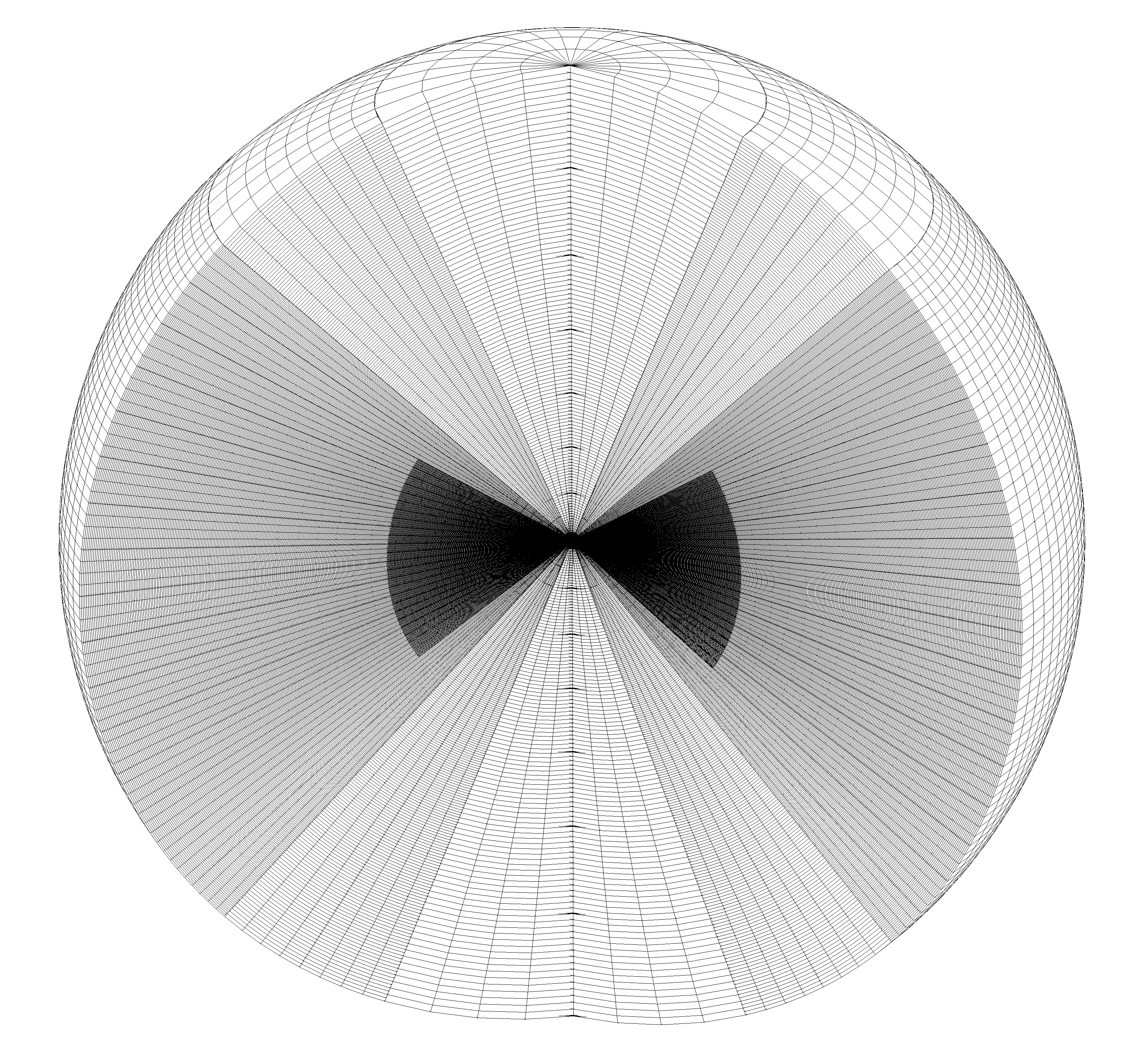}
    \caption{Visualization of the fiducial-resolution simulation mesh in 3D, with a cutout to show internal zones.}
    \label{fig:mesh}
\end{figure}

Our computational domain spans $r\in\left[5r_g,290r_g\right]$, $\theta\in[0,\pi]$, and $\phi\in[0,2\pi)$. Our fiducial pair of simulations have a base resolution of $(N_r,N_{\theta},N_{\phi}) = (112\times32\times32)$ and 3 additional levels of static mesh refinement (SMR), with each level double the resolution of the one below (Figure~\ref{fig:mesh}). This gives us a maximum effective resolution of $(N_r,N_{\theta},N_{\phi}) =(896\times256\times256)$ for $r<100r_g$ and height $\lesssim 2H$, resulting in 20 cells per scale height $H$. Using SMR in this way, our resulting mesh (Figure~\ref{fig:mesh}) is de-resolved near the poles, increasing cell sizes in this region and avoiding untenably small timesteps that would arise from the CFL condition. As a result of this mesh construction we are able to simulate the full spherical domain including the poles. Our two high-resolution runs use a base resolution of $(N_r,N_{\theta},N_{\phi}) = (112\times64\times64)$, i.e. double the resolution in $\theta$ and $\phi$, but are otherwise identical. This gives a maximum effective resolution of $(N_r,N_{\theta},N_{\phi}) =(896\times512\times512)$ and 40 cells per scale height $H$.

The base grid is logarithmically-spaced in $r$ and linearly spaced in $\theta$ and $\phi$. We use ``polar'' boundary conditions (BCs) at the poles \citep[see][for details]{athena++}, periodic BCs in the $\phi$-direction, and outflow BCs on the inner and outer $r$ boundaries. Our polar boundary condition allows magnetic field lines to correctly pass through the coordinate singularity, and allows a polar flux bundle to stay in place without (unphysically) being destroyed by the boundary.

\subsection{Initial Conditions}\label{subsec:initial_cond}
To contrast with previous work and further explore robustness to initial conditions, we initialize our disks with an approximate disk-like configuration following that of \citet{HoggReynolds2018A}, rather than the commonly-used Fishbone-Moncrief torus \citep{FishboneMoncrief1976}.

We initialize our disks with a power-law density profile
\begin{equation}
    \rho(R) = \rho_0 R^{-3/2} e^{-(z^2)/(2c_s^2R^3)}\,F_{\rm s}(R),
\end{equation}
where $R$ is the cylindrical radius, $F_s(R)$ is a function motivated by the zero-torque inner boundary condition of \cite{ShakuraSunyaev1973},
\begin{equation}
    F_s(R) = 1 - \left(\frac{r_{\rm ISCO}}{R}\right)^{1/2}.
\end{equation}
$\rho_0=100$ is a scaling factor set to give ${\rm max}\lbrace\rho(r)\rbrace\sim1$, and $c_s$ is the isothermal sound speed in a thin disk. The resulting density scale-height of the initial disk $H_i$ is given by,
\begin{equation}
    c_s = v_{\rm K} H_i
\end{equation}
for Keplerian rotational velocity $v_K$. 

To this density profile, we add a non-axisymmetric perturbation of the form $\sin(10\phi)$ at the 1\% level. Breaking the formal axisymmetry of the initial conditions lessens the impact of (transitory) channel modes and hastens the development of saturated turbulence.

The initial velocity profile is that of a test-particle on a circular orbit (i.e. ``Keplerian'') in the Pseudo-Newtonian potential,
\begin{equation}
    v_{\phi} = \frac{\sqrt{R}}{R-2},
\end{equation}
with the other components of velocity set initially to zero ($v_r=v_\theta=0$). This initializes the disk close to, but not exactly matching, its natural equilibrium state, reducing the time needed for the disk to settle into a quasi-steady state and the simulation data to become useful.

The initial magnetic field is purely toroidal field, set by
\begin{equation}
    B_{\phi} = \sqrt{\frac{2P}{\beta_0}},
\end{equation}
with all other components equal to zero, i.e. the magnetic pressure is set to be a constant fraction $1/\beta_0$ of the gas pressure $P$. We calculate $P$ from the density before the introduction of the non-axisymmetric perturbations. Thus the corresponding magnetic field is axisymmetric and, being toroidal, is hence divergence free everywhere. We explore two cases of field strength: a stronger field scenario ($\beta_0=5$) and a weaker field scenario ($\beta_0=200$). Combined with the two resolutions that we explore (\ref{subsec:sim_setup}), this defines the four simulations presented in this paper and summarized in Table~\ref{tab:runs}.

\begin{deluxetable*}{lcccrrrrr}
\tablenum{1}
\tablecaption{Simulation parameters}
\tablewidth{0pt}
\tablehead{
\colhead{Simulation ID} & \colhead{Max. Resolution} & \colhead{Duration} & \colhead{ISCO orbits} & \colhead{$\beta_0$} & \colhead{$\llangle\beta\rrangle$} & \colhead{$\llangle Q_{\theta}\rrangle$} & \colhead{$\llangle Q_{\phi}\rrangle$} & \colhead{$\llangle\theta_B\rrangle$}\\
\colhead{} & \colhead{($r\times\theta\times\phi$)} & \colhead{($GM/c^3$)} & \colhead{} & \colhead{} & \colhead{} & \colhead{} & \colhead{}
}
\startdata
\lowbeta & $896\times256\times256$ & $1.67\times10^5$ & 2704 & 5 & $9.39$ & $20.56$ & $63.22$ & $11.85$\\
\highbeta & $896\times256\times256$ & $4.02\times10^5$ & 6535 & 200 & $23.22$ & $9.22$ & $27.38$ & $10.62$\\
\lowbetasup & $896\times512\times512$ & $3.15\times10^4$ & 511 & 5 & $7.87$ & $39.89$ & $106.56$ & $11.14$\\
\highbetasup & $896\times512\times512$ & $1.54\times10^5$ & 2502 & 200 & $20.90$ & $10.75$ & $47.82$ & $10.75$\\
\enddata
\tablecomments{Averages denoted by $\llangle X \rrangle$ are taken over $r\in[5,100]$, $\phi$, and three scale heights in $\theta$, at late times in the disk as defined in Section~\ref{subsec:convergence}.}
\end{deluxetable*}\label{tab:runs}

\subsection{Diagnostics and Convergence} \label{subsec:convergence}

MHD accretion disks are highly turbulent systems and thus, even when they have reached a ``quasi steady state'', will be characterized by strong spatial and temporal fluctuations in all quantities. Useful diagnostics of disk structure necessarily require some averaging process. In this paper, spatial averages are denoted by $\langle X\rangle$; by default, this means averaging over $r < 100r_g$, all $\phi$, and across $\theta={\rm constant}$ shells within $\pm 3$ scale heights of the mid-plane. Quantities denoted by $\llangle X \rrangle$ are additionally averaged over time; by default, we average our weak field simulations (\highbeta\ and \highbetasup) for $t\gtrsim2000T_{\rm ISCO}$ ($t>1.2\times10^5~GM/c^3$) and our strong-field simulations (\lowbeta\ and \lowbetasup) for  $t\gtrsim500T_{\rm ISCO}$ ($t>3.0\times10^4~GM/c^3$). Any deviations from these defaults are indicated by $\langle X\rangle^\star$ or $\llangle X\rrangle^\star$ and details specified within the text. 

In addition to looking at the evolution of quantities already described ($\beta, \rho, v_r$), we calculate and track several additional standard quantities. The mass accretion rate across some closed surface $S$ is 
\begin{equation}\label{eq:mass_accretion}
    \dot{M} = \iint_S \rho \bs{v} \cdot d\bs{S},
\end{equation}
which we calculate across the sphere $r=6$, namely the innermost circular stable orbit (ISCO) of this potential. 

The geometrical scale height $h(r)$ is calculated as in \citet{HoggReynolds2018B}, with the scale height at each radius given by
\begin{equation}\label{eq:scale_height}
    \frac{h(r)}{r} = \Biggl\langle \frac{\int[\theta(r) - \bar{\theta}(r)]^2\rho d\Omega}{\int\rho d\Omega} \Biggr\rangle
\end{equation}
for solid angle $d\Omega = \sin\theta\,{\rm d}\theta\,{\rm d}\phi$ and
\begin{equation}
    \bar{\theta}(r) = \frac{\int\theta(r)\rho d\Omega}{\int\rho d\Omega}.
\end{equation}
The radial bins from equation~\ref{eq:scale_height} are then averaged by weighting them according to the radial width at each shell (as cells are logarithmically spaced in $r$), giving a disk-averaged scale height $H$. To prevent confusion with other parameters within the text, we use $H$ to refer to the disk-averaged geometrical scale height rather than $\langle\frac{h(r)}{r}\rangle$.

With the focus of this paper on the magnetic evolution of the disk, a parameter of key interest is the horizon penetrating magnetic flux,
\begin{equation}\label{eq:flux}
    \Phi_{\rm BH} = \frac{\sqrt{4\pi}}{2}\iint_{S_H} \bs{B}\cdot d\bs{S},
\end{equation}
where $S_H$ is the upper half-sphere near the inner boundary of our domain $(r=6,\theta>0)$. This is then re-normalized by the late-time-averaged accretion rate to $\phi = \Phi_{\rm BH}/(\llangle\dot{M}\rrangle r_gc^2)^{-1/2}$ \citep{TchekhovskoyEA2011}. For a disk scale height $H\sim0.3$ and BH spin $a=0$, the transition from SANE to MAD states is expected to occur for $\phi\sim50$ \citep{TchekhovskoyEA2012}\footnote{We note that $\phi\sim50$ is the saturation value in code or Gaussian units. When using Lorentz-Heaviside units the factor of $\sqrt{4\pi}$ in equation~\ref{eq:flux} disappears, giving $\phi\sim15$ instead.}.

We also calculate the Shakura-Sunyaev alpha \citep{ShakuraSunyaev1973},
\begin{equation}\label{eq:alpha}
    \alpha_{\rm SS} = \frac{T_{\hat{r}\hat{\phi}}}{P} = \frac{T_{{\rm Max},\hat{r}\hat{\phi}} + T_{{\rm Rey},\hat{r}\hat{\phi}}}{P}
\end{equation}
where $T_{\hat{r}\hat{\phi}}$ is the fluid frame stress, and $T_{{\rm Max},\hat{r}\hat{\phi}}=-B_r B_{\phi}$ and $T_{{\rm Rey},\hat{r}\hat{\phi}}=\rho u_r \delta u_{\phi}$ are the Maxwell and Reynolds components of that stress, where $\delta u_{\phi}\equiv u_{\phi} - r\Omega_{\rm Kep}$ is the fluctuating component of the $\phi$-velocity. For thin disks, $\alpha_{\rm SS}\sim0.1$, however this value is known to vary with disk thickness, initial magnetic field morphology/strength, and radius within the disk \citep[e.g.][]{PennaEA2013}.

A key question is the degree to which a numerical model of an accretion disk is converged. There are two distinct senses in which we must consider convergence. Firstly, since our disks are initialized in a laminar state that is close to hydrodynamic equilibrium but strongly unstable to the MRI, they must first undergo evolution before settling into a (turbulent) statistical steady state. Only then can the simulation data be considered physically meaningful (out to the radius where we achieve inflow equilibrium). Diagnosing this state informs the times on which we focus our analysis. Secondly, as we are interested in studying the behavior of the magnetic field we need to be sure that the underlying physical mechanisms that govern its evolution are spatially well-resolved within the disk. To assess these criteria, we turn to two measures of convergence: the quality factors $Q_{\theta}$ and $Q_{\phi}$, and the magnetic tilt angle $\theta_B$. Additionally, we are able to do a very limited resolution study by comparing our low and high resolution simulations. 

The quality factors, $Q$, defined by \citet{NobleEA2010} as
\begin{align}
    Q_{\theta} &= \frac{\lambda_{{\rm MRI},\theta}}{r\Delta\theta} \\
    Q_{\phi} &= \frac{\lambda_{{\rm MRI},\phi}}{R\Delta\phi},
\end{align}
are a direct measure of the resolvability of the linear MRI modes via a comparison of the characteristic wavelength of the MRI modes $\lambda_{\rm MRI}=2\pi v_A/\Omega$, where $v_A$ is the Alfv\'{e}n velocity, with the size of the basic simulation voxel. The $Q$'s are simply a measure of the number of resolution elements that span one wavelength of the fastest growing MRI mode.

Values as low as $Q_{\theta}=6-8$ are sufficient to capture the linear MRI \citep{FlockEA2010} in low-spatial order methods, but must rise to $Q_{\theta}>10$ with $Q_{\phi}>20$ to capture the non-linear interactions between MRI modes \citep{HawleyEA2011,HawleyEA2013,SorathiaEA2012}, although a higher $Q_{\phi}$ or $Q_{z}$ (when in cylindrical coordinates) may offset a slightly lower $Q_{\theta}$.

Our second diagnostic tool is the magnetic tilt angle, $\theta_B$. First used by \cite{GuanEA2009}, $\theta_B$ is a measure of the correlations between the radial and toroidal fields,
\begin{equation}
    \theta_B = -\arctan\left<\frac{B_r}{B_\phi}\right>.
\end{equation}
\cite{GuanEA2009} originally found this to have a characteristic value of $\theta_B\sim15^\circ$ for a well-resolved saturated MRI-driven turbulence with later works on local \citep{HawleyEA2011} and global \citep{SorathiaEA2012,HawleyEA2013,HoggReynolds2016,HoggReynolds2018A} simulations revising this number down to 11-13$^\circ$. The physics dictating the saturation of MRI-driven turbulence and hence the value of the tilt angle is still not clear, although \citet{Pessah2010} describe such behavior with a model in which saturation corresponds to disruption of linear-MRI modes by the parasitic instabilities of \citet{GoodmanXu1994}. Our primary focus with $\theta_B$ is not the final value itself, as this is not a direct measure of resolution and so may be affected by other factors in the disk, but we instead look for a stabilization in the value of this parameter.

\begin{figure}[t]
    \plotone{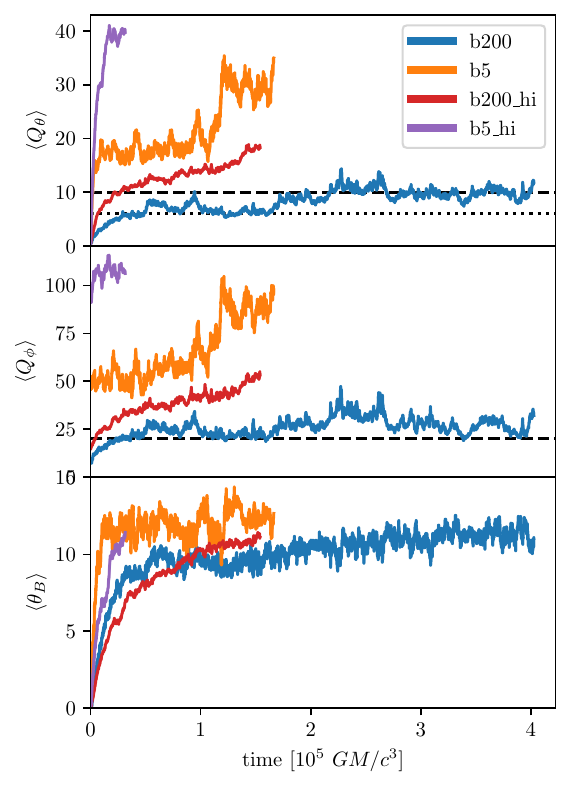}
    \caption{Convergence metrics: ({\it top} and {\it middle}) quality factors in the $\theta$- and $\phi$-directions, and (bottom) magnetic tilt angle, $\theta_B$. The dotted line indicates the minimum value required to resolve the linear MRI, while the dashed lines represent the requirements for resolving the non-linear MRI. }
    \label{fig:qual}
\end{figure}

\begin{figure*}[t!]
    \plotone{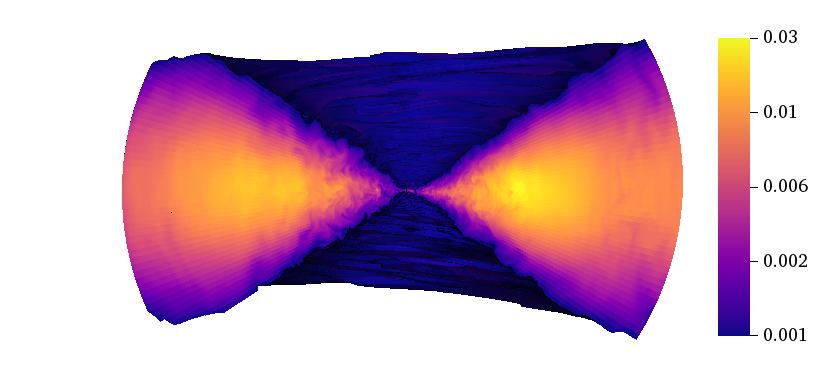}
    \caption{Three-dimensional side-on volume rendering of the simulation domain for one of our fiducial runs, \highbeta, at a late time in the disk. Color denotes density.}
    \label{fig:pretty_picture}
\end{figure*}

The time-evolution of these convergence metrics for all of our runs are shown in Figure~\ref{fig:qual}, and the late-time average is reported in Table~\ref{tab:runs}. Through these metrics we find that our strong-field simulation \lowbeta\ is well-resolved with $Q_{\theta}\sim16$, $Q_{\phi}\sim38$, and a tilt angle that stabilizes at $\theta_B\approx12^\circ$. The numbers are less comfortable for \highbeta\ (Figure~\ref{fig:qual}). The lower values of $Q_{\theta}\sim9$ indicates that \highbeta\ only marginally resolves the dominant linear MRI modes, although $\theta_B$ does eventually stabilize for $t\gtrsim2\times10^5~GM/c^3$, albeit at a notably lower value ($\theta_B\sim10)$ than the stronger field case.

A more direct way to assess the effects of resolution is to compare our standard resolution runs (\highbeta\ and \lowbeta) with their high-resolution counterparts (\highbetasup\ and \lowbetasup). As expected, these high-resolution runs achieve much higher $Q_{\theta}$ and $Q_{\phi}$ values, well above the fiducial levels needed. The magnetic tilt angle $\theta_B$ take longer to stabilize than in their lower-resolution counterparts but eventually stabilize at comparable levels to the low-resolution runs.

We conclude that all simulations have achieved a converged state by the end of their respective runs, although \highbeta\ may be only marginally resolving both the linear and non-linear evolution of the MRI.

\section{Simulation Results}\label{sec:results}

\subsection{Disk structure and evolution}\label{subsec:evolution}

\begin{figure}[t!]
    \plotone{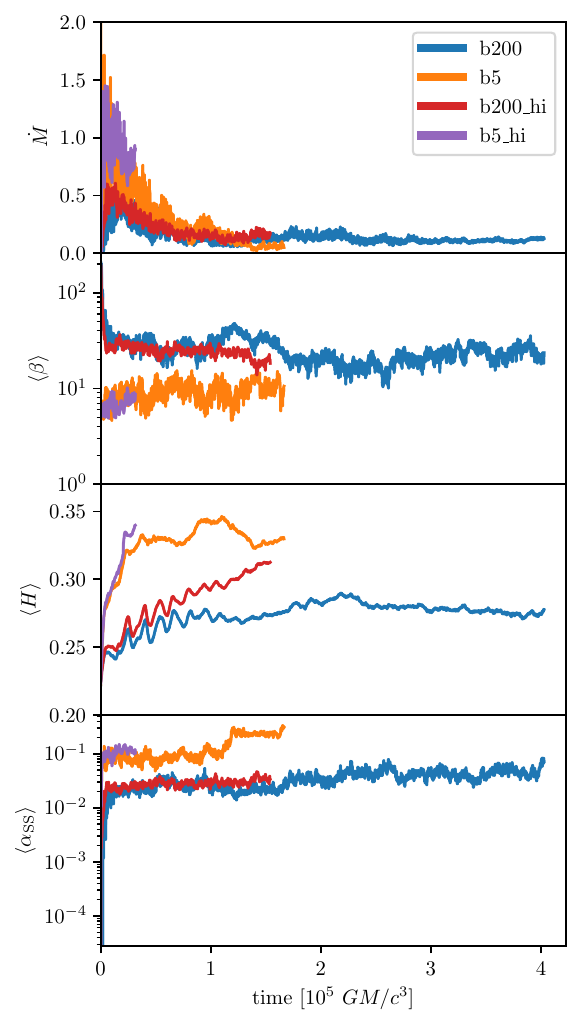}
    \caption{({\it Top}) Mass flux $\dot{M}$ through the surface at $r=6r_g$, ({\it top middle}) spatially-averaged magnetic plasma $\langle\beta\rangle$, ({\it bottom middle}) spatially-averaged geometric scale height $\langle H\rangle$, and ({\it bottom}) spatially-averaged Shakura-Sunyaev alpha $\langle \alpha_{\rm SS}\rangle$.}
    \label{fig:characteristics}
\end{figure}

Starting from the initial state described in Section~\ref{subsec:initial_cond}, the disks undergo rapid evolution driven principally by the growth and eventual saturation of MRI-driven turbulence. The outer parts of the simulated disks never achieve a state of inflow equilibrium; we focus our analysis on the inner (and most highly resolved) regions of the disks within $r=100r_g$. Figure~\ref{fig:pretty_picture} shows a late-time three-dimensional volume rendering of run \highbeta, and Figure~\ref{fig:characteristics} shows the evolution of mass accretion rate across the ISCO as well as the disk-averaged plasma-$\beta$, geometric scale height $H$, and Shakura-Sunyaev $\alpha_{\rm SS}$. Both \lowbeta\ and \highbeta\ show a characteristic early peak and slow drop-off in mass accretion rate $\dot{M}$, as expected for disks with finite mass (Figure~\ref{fig:characteristics}). \highbeta\ stabilizes to a higher mass accretion rate of $\dot{M}\sim0.25$, possibly related to the presence of a density lump that inhibits early-time accretion. In all models, we find that $\langle\beta\rangle$ reaches an approximately steady value relatively quickly. The memory of the strength of the initial magnetic field is retained, with run \highbeta\ stabilizing at $\langle\beta\rangle\sim 20$ and run \lowbeta\ stabilizing at $\langle\beta\rangle\sim 10$. The initial magnetization also makes a difference to the disk thickness, with stabilization at $\langle H\rangle\sim0.27$ and $\langle H\rangle\sim0.32$ for \highbeta\ and \lowbeta\ respectively. The dominant pressure dictating the disk thickness in all cases is gas pressure, with a modest amount of additional magnetic support thickening the disk in the stronger initial field case \lowbeta. The spatially-averaged $\langle\alpha_{\rm SS}\rangle$ (lower panel) remains relatively stable around $10^{-2}$-$10^{-1}$, as is expected for accretion disks, with a small jump in \lowbetasup\ at late times which coincides with a thinning of the disk (i.e. lower $H$). 

For the stronger field case, doubling the $\theta$- and $\phi$-resolution (going from \lowbeta\ to \lowbetasup) makes very little difference to the overall evolution of the disk, with $\dot{M}$, $\langle\beta\rangle$ and $\langle H\rangle$ following very similar time-traces. In the case of the weaker field model, the early-to-intermediate timescale evolution is very similar for the high-resolution models whereas the late-time magnetization is a little higher ($\langle\beta\rangle$ marginally smaller) leading to a slightly thicker disk. This further suggests that the fiducial resolution weaker field run \highbeta\ only marginally resolves the non-linear evolution of the MRI.

\begin{figure}[t!]
    \plotone{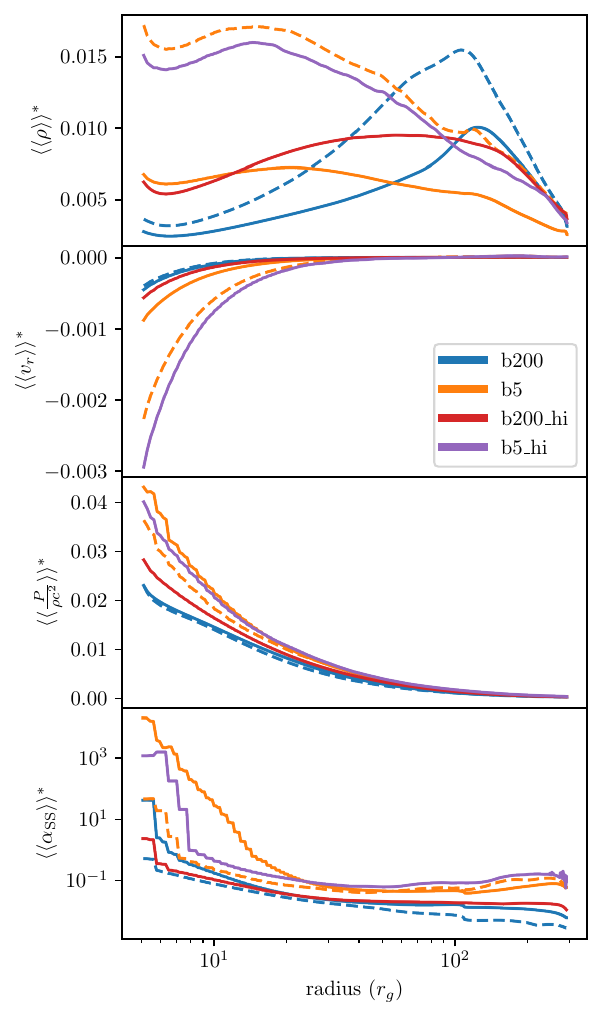}
    \caption{Late-time-averaged profiles for the density ({\it top}), radial velocity ({\it top middle}), temperature ({\it bottom middle}), and Shakura-Sunyaev alpha ({\it bottom}) as a function of disk radius. Dashed lines for \highbeta\ and \lowbeta\ are averaged over the same time windows as \highbetasup\ and \lowbetasup\ respectively. Note: Spatial averages are taken azimuthally and within $3H$.}
    \label{fig:profiles}
\end{figure}

The radial profiles of density $\langle\langle\rho\rangle\rangle^*$, radial velocity $\langle\langle v_r\rangle\rangle^*$, temperature $\langle\langle P/\rho c^2\rangle\rangle^*$, and effective Shakura-Sunyaev $\langle\langle \alpha_{\rm SS} \rangle\rangle^*$ are shown for all four runs in Figure~\ref{fig:profiles}. These are time-averaged from the point where the run is deemed to have stabilized (as laid out in Section~\ref{subsec:convergence}) out to the end of the simulation. Given that the different runs have quite different end-times, some caution must be exercised in making direct comparisons between profiles from different runs. To aid in this comparison, we also provide the profiles for the low-resolution runs (\highbeta\ and \lowbeta) averaged over the same time range as the high-resolution counterparts shown as dashed lines. For example, the large difference in the overall normalization of density between \lowbeta\ and \lowbetasup\ is almost all due to the fact that the lower-resolution run was evolved much longer and hence drained much more of its mass; when the average is calculated over the same time range, this difference largely disappears. However, the comparison of \highbetasup\ and \lowbeta\ is particularly interesting. Both runs well-resolve the MRI and have evolved for a similar time, with the principal difference being the initial magnetic field strength. We find that both runs have very similar radial velocity profiles during these late-time averages, although \lowbeta\ experienced high-accretion rate transients at early times that suppresses its density. This early-time behavior is clearly seen in the profiles for \lowbetasup\ which clearly shows significantly higher radial inflow within the central 30$r_g$. 

\label{subsec:hydro_differences}
\begin{figure*}[t!]
    \plotone{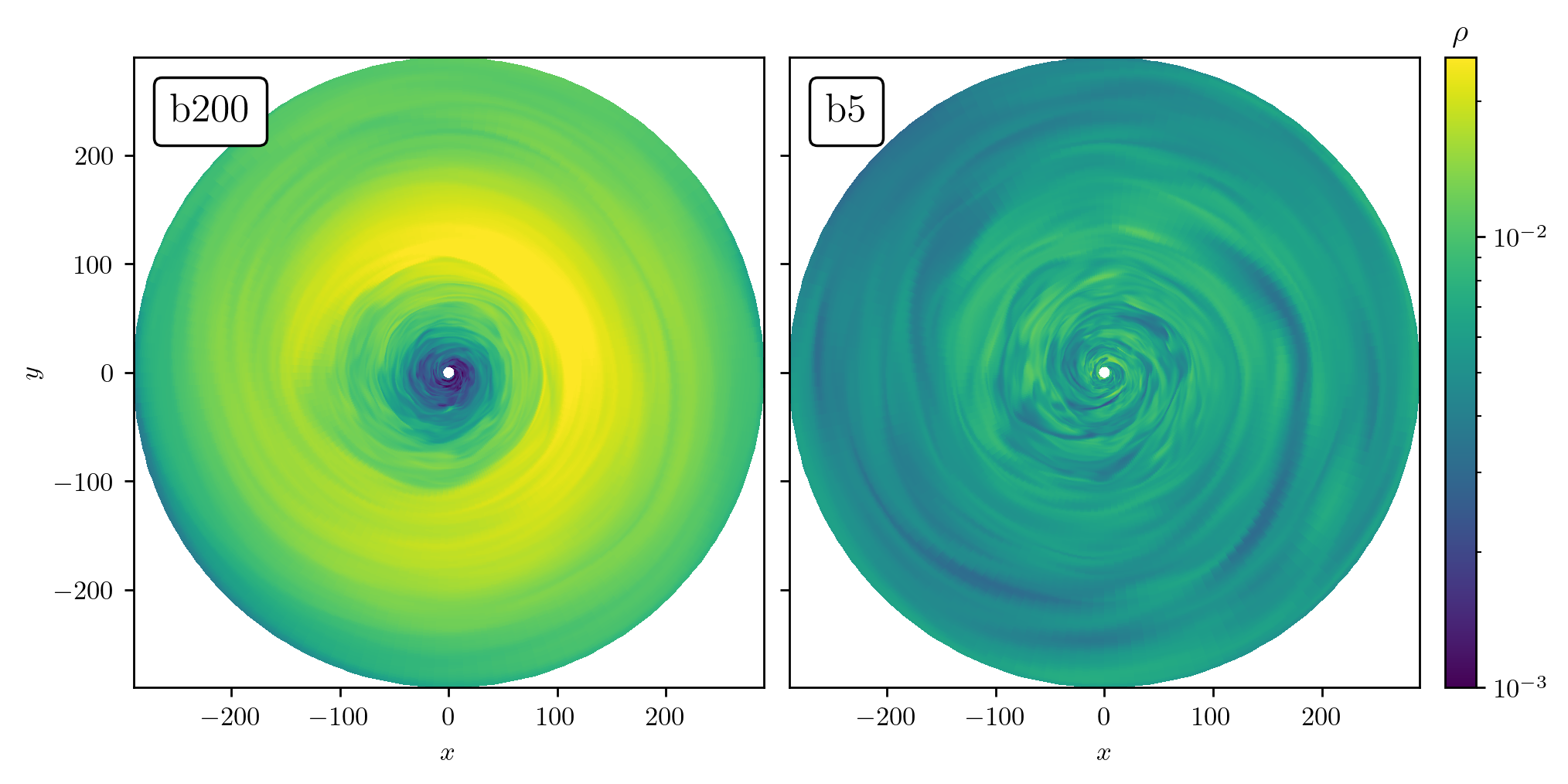}
    \caption{Late-time snapshots of midplane density for \highbeta\ ({\it left}) and \lowbeta\ ({\it right}), taken at $t=4.05\times10^5~GM/c^3$ and $t=1.25\times10^5~GM/c^3$ respectively.}
    \label{fig:dens_midplane}
\end{figure*}

\begin{figure*}[t!]
    \epsscale{2.0}
    \plottwo{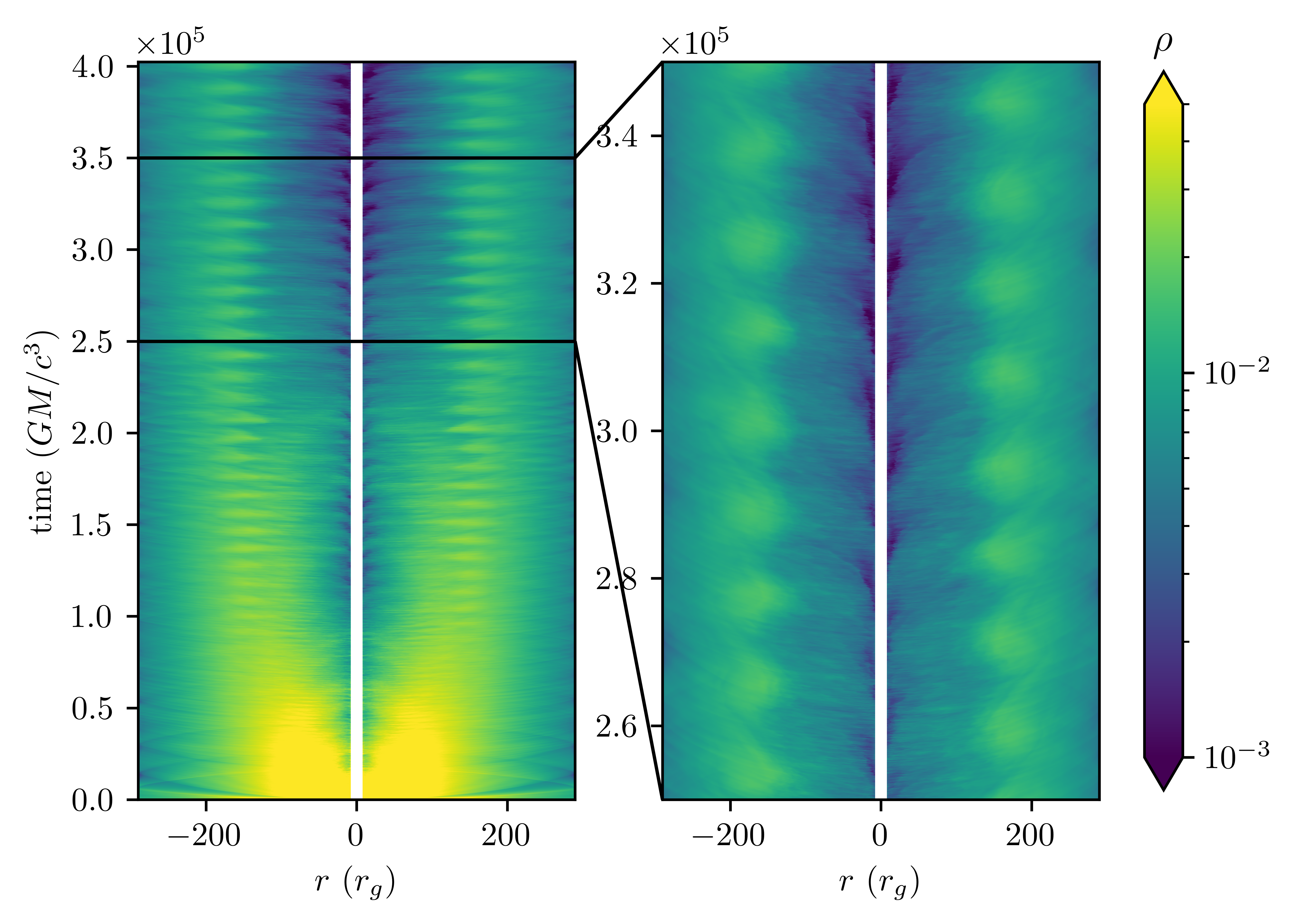}{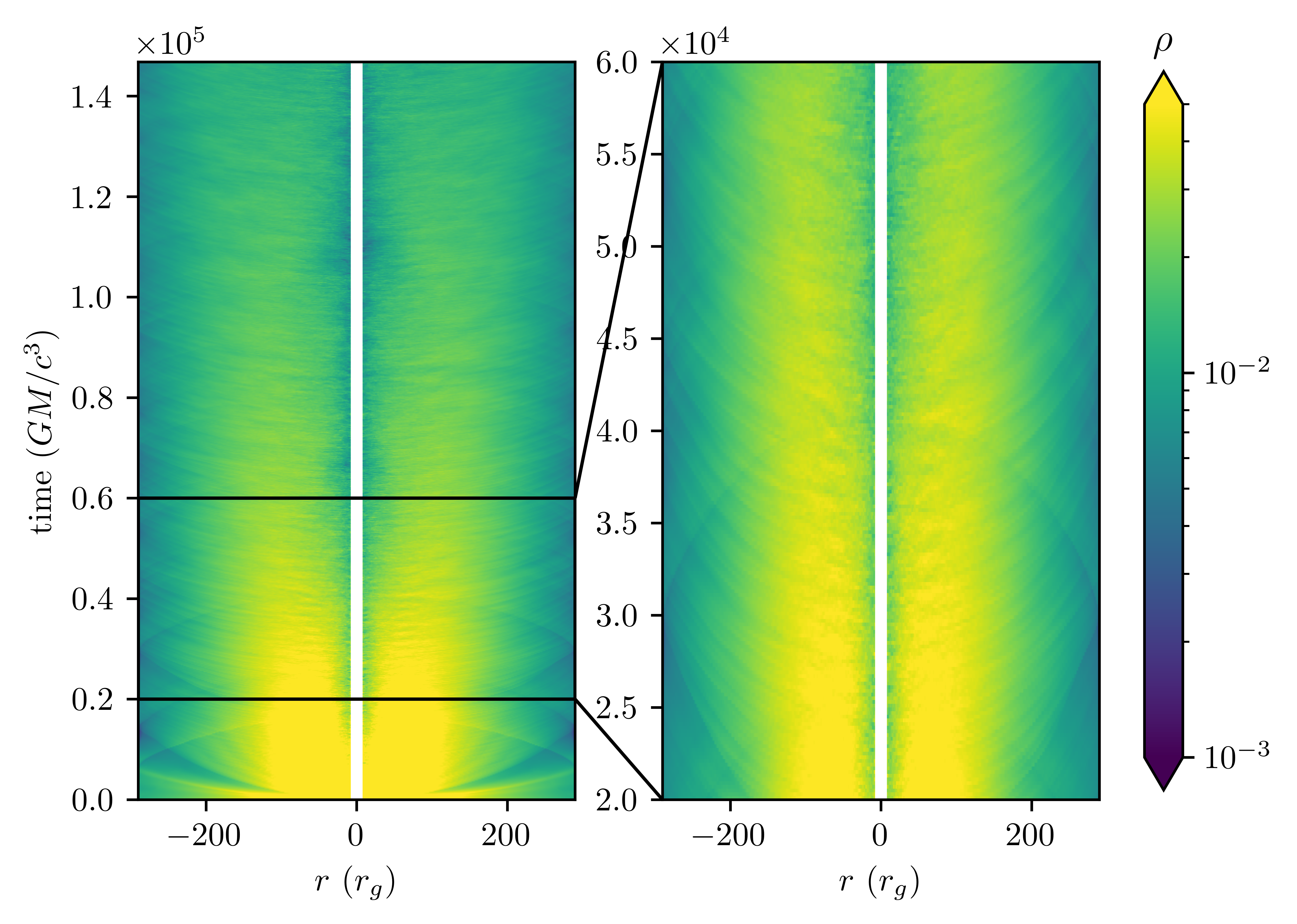}
    \caption{Density along a midplane slice ($\theta=\pi/2$, $\phi=\lbrace0,\pi\rbrace$) over time, for \highbeta\ ({\it top}) and \highbetasup\ ({\it bottom}). The overdensity, or $m=1$ mode, is clearly visible in \highbeta\ as striations between radii $100-200r_g$.}
    \label{fig:lump}
\end{figure*}

The density profile of \highbeta\ shows a broad hump around $r=100r_g$. This signposts a particularly interesting feature of this disk that, we argue, is a subtle and pernicious imprint of the SMR de-refinement boundaries. The overdensity, or $m=1$ mode, develops at a relatively early time of $t\sim1\times10^5~GM/c^3$ and is then sustained throughout the subsequent evolution. This structure is most apparent in the mid-plane density map shown in Figure~\ref{fig:dens_midplane} (left panel). The time-dependence of this lump can be seen by forming space-time plots of the density along a midplane line that cuts through the center, $(\theta=\pi/2,\, \phi=0,\pi)$; see Figure~\ref{fig:lump} (top panel). The sloshing motion of this lump as it orbits the central body is seen as the long-lived ``see-saw'' feature, causing an excavation of material from the central region and creating the density maximum at $r\sim100r_g$ (Figure~\ref{fig:profiles}). An examination of the mid-plane velocity field reveals this lump to be a large scale vortex. It has been argued that such features might arise from a Rossby Wave Instability \citep[RWI;][]{LovelaceEA1999} or generalized form of the Papaloizou-Pringle Instability \citep[PPI;][]{PapaloizouPringle1984}. The presence of the RWI within disks has already been argued in the case of Tidal Disruption Events (TDEs) and the spirals it produces may explain the observed flaring behaviors in Sgr A* \citep{FalangaEA2007,Mignon-RisseEA2021}. A similar feature was also reported in an MHD simulation by \citet{MishraEA2020}, who observed spiral-like inhomogeneities in the density of disks with initially weak poloidal fields, however these simulations were only run for 23 orbits (versus our $\sim6500$ for \highbeta) and so the late-time evolution of the effect was not noted.

\begin{figure*}[t!]
    \plotone{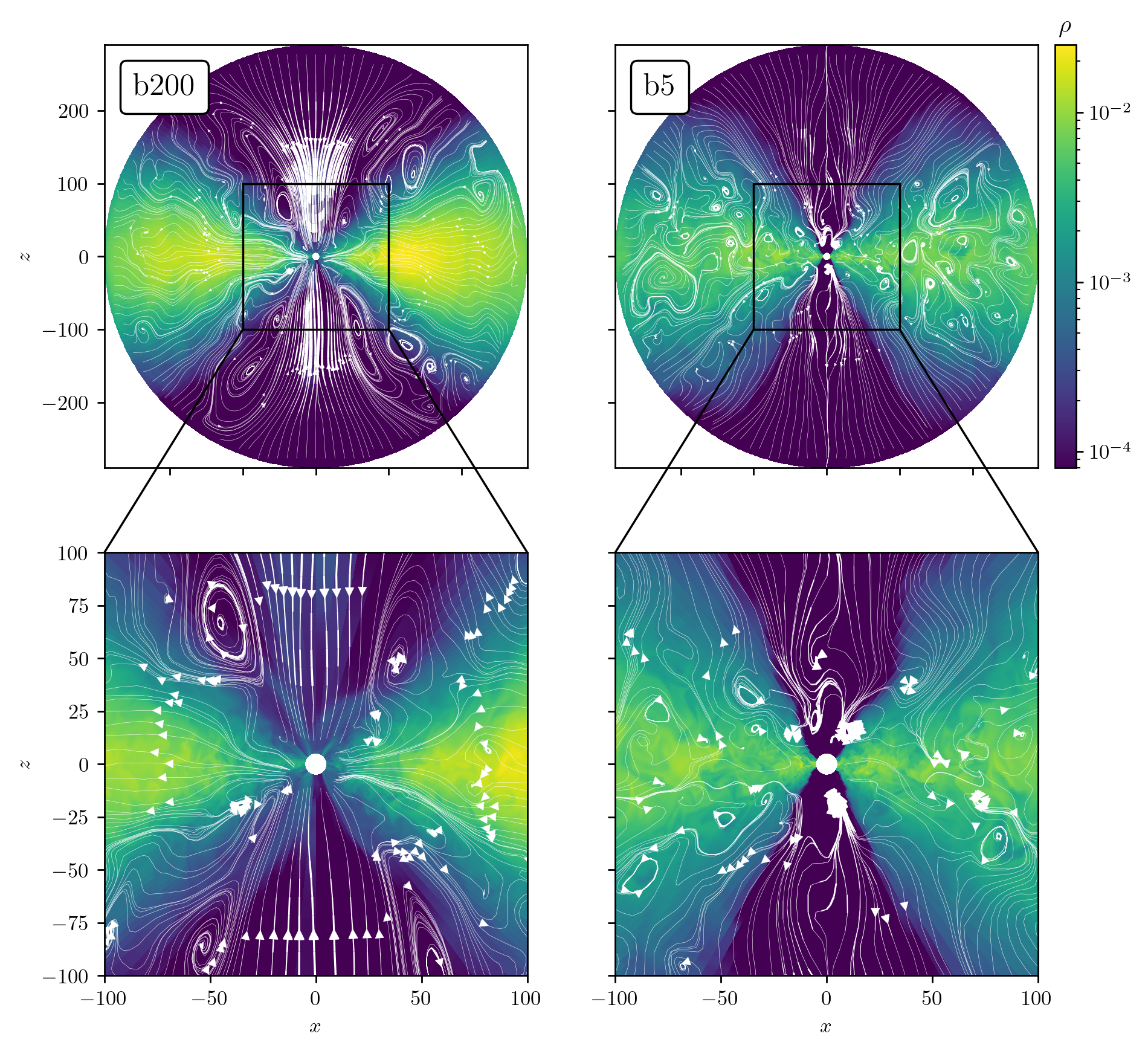}
    \caption{Velocity streamlines for the whole domain ({\it top}) and zoom-in on the $r<100r_g$ region ({\it bottom}) overlaid onto density for \highbeta\ ({\it left}) and \lowbeta\ ({\it right}). Relative velocity magnitude is indicated by the linewidth of streamlines. Stills are taken from late times: $t=4.05\times10^5~GM/c^3$ for \highbeta\ and $t=1.25\times10^5~GM/c^3$ for \lowbeta.}
    \label{fig:v_stream}
\end{figure*}

As seen in \citet{MishraEA2020}, the lump creates a kind of anomalous accretion where material preferentially accretes from the atmosphere or ceiling of the disk rather than in the midplane. This is clearly seen in Figure~\ref{fig:v_stream} which shows the poloidal velocity field on a vertical cross-sectional slice for late-time snapshots from \highbeta\ and \lowbeta. Run \highbeta's velocity streamlines (Figure~\ref{fig:v_stream}, left panel) show strong inflows in the jet regions of the \highbeta\ simulation and eddies along the disk-jet boundary which take material out of the disk and deposit it in the jet streams, while the radial velocity within the disk is reduced (Figure~\ref{fig:profiles}).

The \lowbeta\ disk, by comparison, does not develop such an overdensity and so retains a significant amount of its mass below its $\rho_{\rm max}$ at $20r_g$ (Figures~\ref{fig:profiles} and \ref{fig:dens_midplane}, right panel). This smoother disk is also accompanied by a greater inflow velocity $v_r$ and a higher temperature $P/\rho c^2$. Thus stronger magnetic fields suppress the formation of the vortex. More importantly, an examination of the corresponding space-time diagram (Figure~\ref{fig:lump}, bottom panel) shows that the lump is not observed in the high-resolution simulation \highbetasup. This fact together with the location of the \highbeta\ lump, strongly suggests that the lump is an artifact caused by vorticity generation at the SMR refinement boundary which affects only the weaker-field case. This is a cautionary note for any long-duration accretion disk models that employ SMR.

\subsection{Magnetic Field Strength, Morphology, and Evolution}

A primary motivation for our study is to understand the evolution of the magnetic field, particularly the growth and morphology of the poloidal field, in these disks that start with purely toroidal configurations. Furthermore, we wish to understand whether the strength of the initial toroidal field has any long lasting imprint on the magnetic evolution of the disks.

\begin{figure*}[ht!]
    \plotone{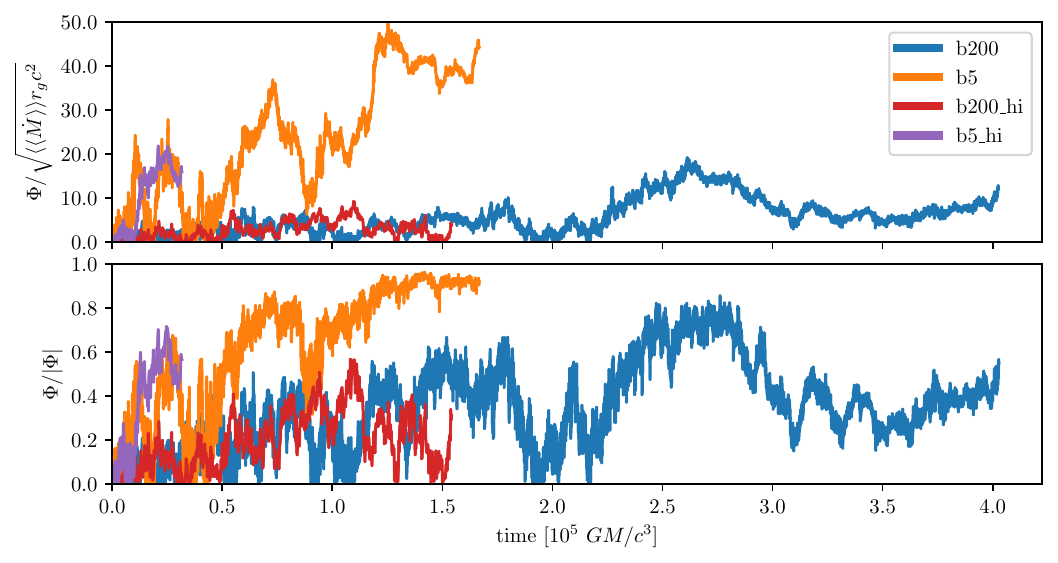}
    \caption{({\it Top}) Magnetic flux, $\Phi_B = \int{B_r\cdot dS}$, through the ISCO ($r=6r_g$). The signed flux is calculated for each hemisphere individually before the absolute value is taken and the two hemispheres averaged to give $|\langle\Phi_B\rangle|$. The values are re-normalized according to \citet{TchekhovskoyEA2011} (Section~\ref{sec:results}). ({\it Bottom}) Ratio of signed and unsigned magnetic fluxes, indicating how much of the flux is aligned in a common direction. }
    \label{fig:flux}
\end{figure*}

Our initially weak-field simulation, \highbeta, shows rapid growth of the magnetic field at early times, settling to $\langle\beta\rangle\sim20$ (Figure~\ref{fig:characteristics}), compared to $\langle\beta\rangle\sim10$ for the initially strong-field case \lowbeta. Interestingly, we see the development of large-scale poloidal field in both cases. Figure~\ref{fig:flux} shows the horizon-threading (hence necessarily poloidal) flux as defined in equation~\ref{eq:flux}. While there is clearly non-zero poloidal flux in all cases, \highbeta\ shows a consistently low flux $\phi\sim5$-$15$ with little sustained growth, whereas \lowbeta\ exceeds $\phi=40$ and nears the transitional value for a MAD state of $\phi=50$ \citep{TchekhovskoyEA2012, LiskaEA2020} for a non-spinning BH, although fails to cross. The high-resolution simulations \lowbetasup\ and \highbetasup\ again match their lower-resolution analogues remarkably well.

It is not merely the net horizon-crossing flux that matters, though, when considering features like jet launching. The degree of order of this field as it crosses the inner boundary is also important to questions such as whether the jet is ``striped'' \citep{GianniosUzdensky2019}. Most disk models that start with  initially-poloidal field configurations end up with a well-organized (unidirectional) horizon-threading flux bundle. However, that is not necessarily the case for our models which begin with initially-toroidal field. To examine this, we calculate the ratio of signed to unsigned fluxes threading the inner boundary of our simulation,
\begin{equation}
    \phi_{\parallel} = \frac{\iint_S \bs{B}\cdot d\bs{S}}{\iint_S |\bs{B}|\cdot d\bs{S}},
\end{equation}
where $S$ is a hemisphere surface at $r=6r_g$ and the final value given is the average of the two hemispheres. A value of $\phi_{\parallel}=1$ indicates agreement in the direction of the field, while a value $\phi_{\parallel}=0$ indicates that (for example) for every positive $B_r$ there is an equal negative, canceling the flux everywhere on the surface integrated.

With this new measure, we see that \lowbeta\ has a high degree of field order with $\phi_{\parallel}\sim1$ at late times, while \highbeta\ barely reaches $\phi_{\parallel}\sim0.8$ at $t\sim2.5\times10^5$ before decreasing again and spending much of its time with $\phi_{\parallel}<0.6$. Again, we see a high degree of agreement between the high- and low-resolution runs.

\begin{figure*}[t!]
    \plotone{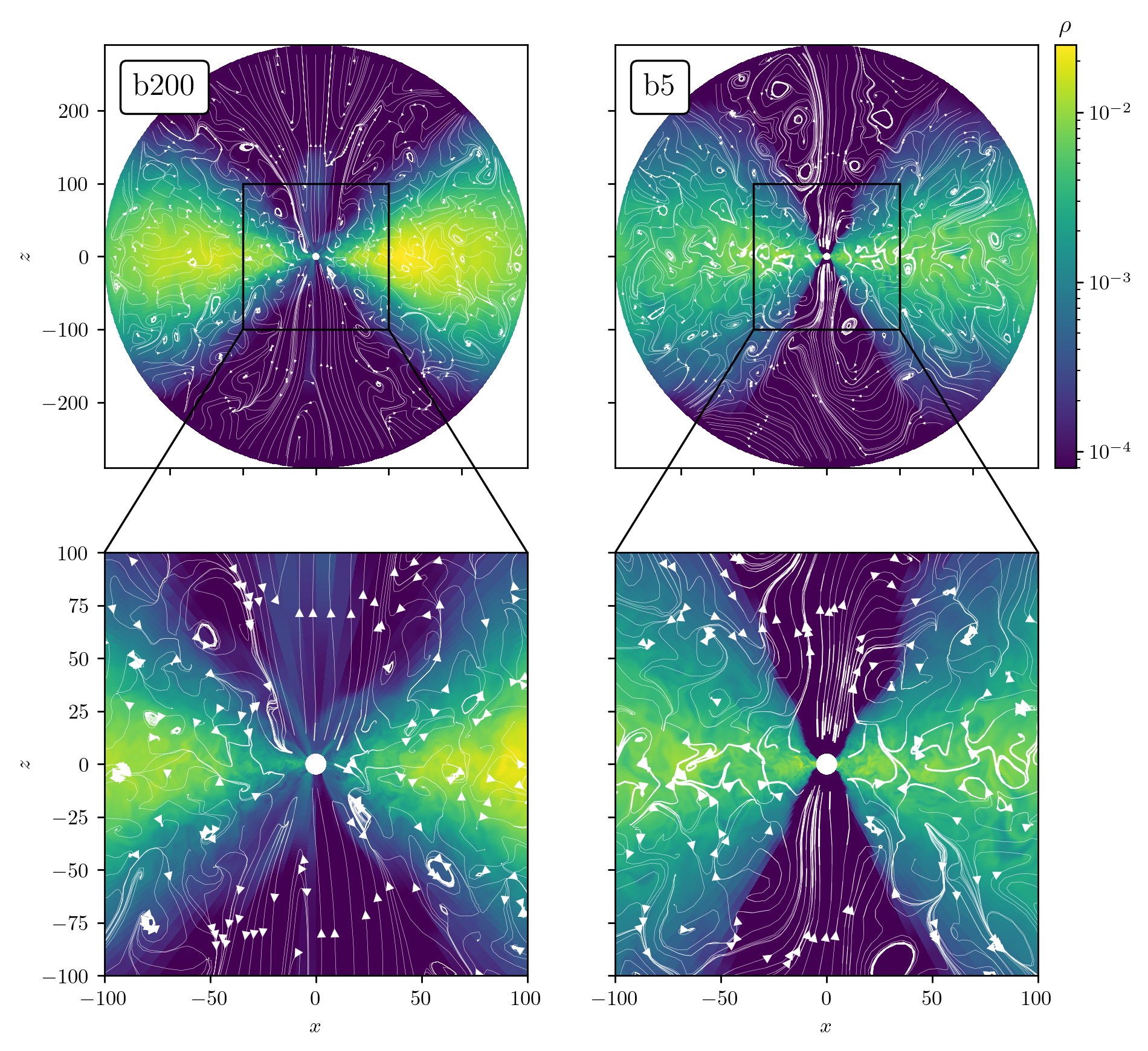}
    \caption{As in Figure~\ref{fig:v_stream}, but with magnetic field streamlines.}
    \label{fig:bcc_stream}
\end{figure*}

These nature of these differences are seen more clearly in renderings of the poloidal magnetic field morphology at late times in Figure~\ref{fig:bcc_stream}. In \highbeta, close to the black hole, we see a well-defined field direction dominating the polar regions and then a strong region of oppositely directed field along a lower-latitude boundary funnel. The mid-plane field is weak and disorganized, existing within small magnetic loops distributed seemingly at random. Recalling that this region is flowing {\it inwards}, some of this magnetic topology may be due to large-scale circulations between the disk atmosphere and the polar regions. By contrast, \lowbeta\ displays noticeably stronger fields in the midplane of the disk, and where there are magnetic loops these are larger than in \highbeta. The polar region contains an organized and sustained flux bundle along which there is outflowing material. It must be recognized, though, that while $\phi_\parallel\sim1$ for this case, the flux bundle still contains interesting structure (field loops and current sheets) downstream, possibly due to Kelvin-Helmholtz modes excited at the jet-disk boundary. Such large magnetic field loops, particularly those on the right side of the upper jet region and the left side of the lower jet region of Figure~\ref{fig:bcc_stream}, are reminiscent of the large-scale poloidal field loops identified by \citet{LiskaEA2020}, and so leads us to tentatively conclude that simulation \lowbeta\ is capable of generating a large-scale poloidal field from an initially-strong toroidal field, however evidence is much weaker for the weak-field case \highbeta.

\begin{figure*}[ht!]
    \plotone{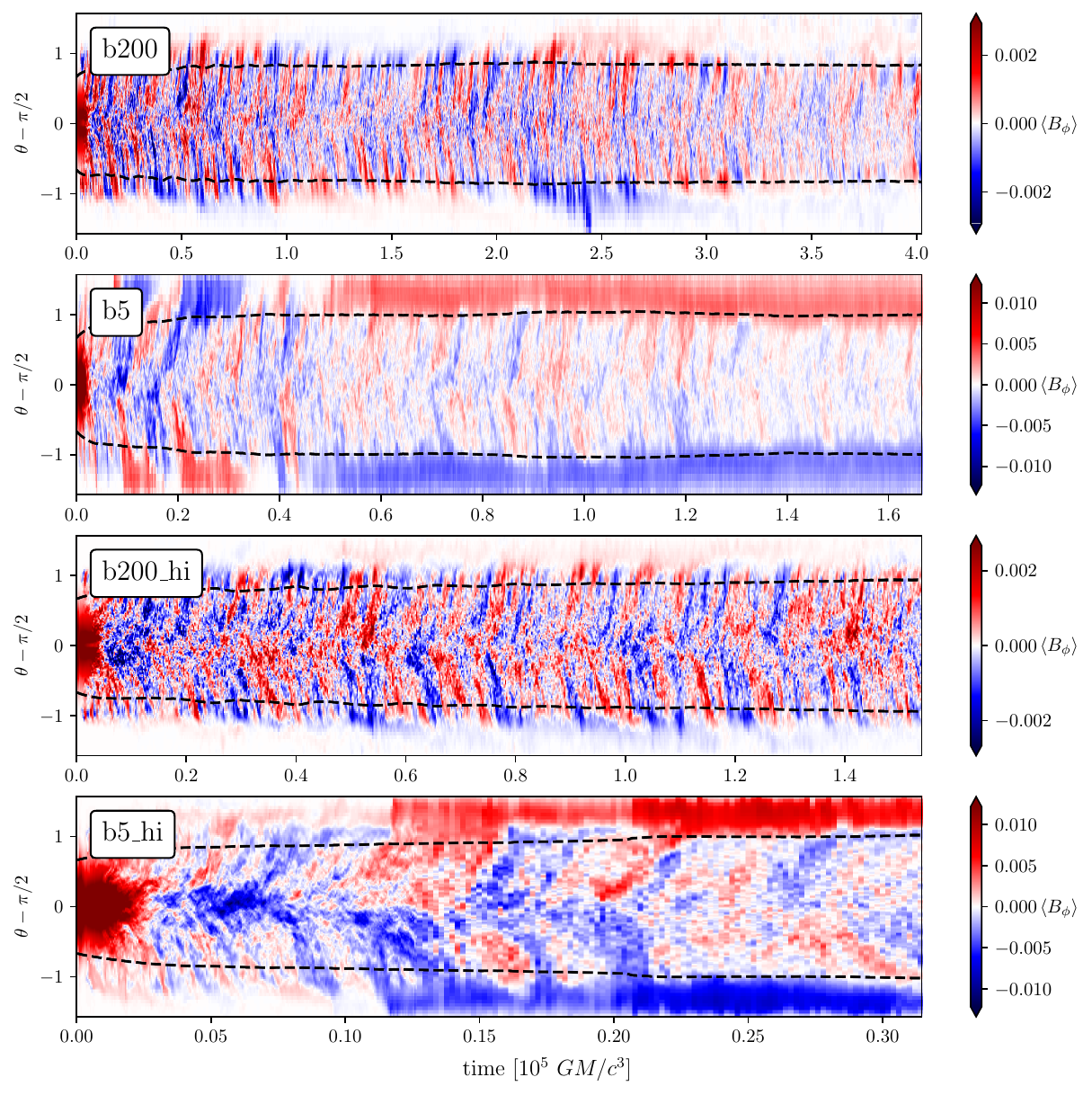}
    \caption{Butterfly diagrams of the azimuthally-averaged $B_{\phi}$ at $r=25r_g$ showing signs of a sporadic butterfly pattern. Dashed lines mark $\pm 3H$. Note there is a change in output cadence for \lowbetasup\ ({\it bottom}) for $t>1.25\times10^4~GM/c^3$.}
    \label{fig:butterfly}
\end{figure*}

Our discussion of magnetic evolution so far has focused on the poloidal field, but we note that there is also interesting dynamics in the toroidal field. We recover the well-known butterfly patterns -- a ``feathering'' pattern in the space-time diagram for the azimuthally-averaged $B_{\phi}$ at a particular radius (Figure~\ref{fig:butterfly}). These are interpreted as dynamo-cycles, an emergent phenomenon that arises from the non-linear MRI physics \citep{Rincon2007,GresselPessah2015,HoggReynolds2018B}. In \highbeta\ (Figure~\ref{fig:butterfly}, top panel), the butterfly pattern is rapidly established and is then sustained for the full duration of the simulation. This plot also explicitly shows that (after azimuthal averaging) a non-zero but weak toroidal flux exists within the polar region. These findings completely track over into the higher-resolution version of this model \highbetasup. By contrast, \lowbeta\ establishes a dominant toroidal field sign in the polar regions at a relatively early time ($t\sim5\times10^4~GM/c^3$), coincident with the creation of the strong and organized poloidal flux bundle (Figure~\ref{fig:flux}). From this time onward the butterfly pattern is significantly weakened, showing that the magnetic configuration in the polar region is affecting the operation of the dynamo cycle.

While we have attributed its presence to the (artificial) SMR refinement boundary, it is still interesting to consider the effect of the $m=1$ lump found in \highbeta\ and discussed in Section~\ref{subsec:evolution}. As the inflow of material from large radius is impeded by the lump, a lower-density cavity forms within the inner regions of the disk (Figure~\ref{fig:profiles}). This increases the magnetization of the region (decreases the plasma-$\beta$) which may be responsible for weakening the amplitude of the dynamo cycles in \highbeta\ from $t\approx 1.0\times 10^5~GM/c^3$ onward.

\section{Discussion and conclusion} \label{sec:discussion}

In this work we have presented the results of two 3D MHD accretion disk simulations with initially toroidal fields. These simulations were identical in all respects bar their initial field strength, defined through the initial magnetic plasma beta, $\beta = P_{\rm gas}/P_{\rm mag}$. One simulation, \highbeta, was initialized with $\beta=200$, while the other, \lowbeta, was initialized with $\beta=5$. Additionally, two more high-resolution runs were done with double the $\theta$ and $\phi$ resolution for comparison (\highbetasup\ and \lowbetasup; for details, see Section~\ref{sec:model}). 

We find that our two fiducial simulations diverge early in their evolution and remain on separate paths throughout each run, with notable differences in both hydrodynamic and magnetic quantities. We show that these differences (with one exception) are not sensitive to resolution for the values studied. As the only difference between these two simulations is the initial field strength, we conclude that it is this initial magnetization that determines the long-term growth (or lack thereof) of the magnetic field. We produce only a weak large-scale poloidal field in \highbeta\ despite evolving for a very long time ($4.0\times 10^5~GM/c^3$). We do, qualitatively, produce a large-scale poloidal field in \lowbeta, and broadly reproduce the results of \citet{LiskaEA2020} whose simulation features a comparable strong toroidal field. This finding shows that the results of \citet{LiskaEA2020} can be reproduced in a pseudo-Newtonian framework and do not rely on the inclusion of General Relativistic physics or the peculiarities of spinning black holes.

All simulations produced clear--although disordered--butterfly patterns in the sign of $\langle B_{\phi}\rangle$, indicating quasi-periodic cycles of the MRI dynamo. We additionally find that {\itshape all} simulations, both weak and strong fields, establish a dominant signed field in the jet region, however, in \highbeta\ and \highbetasup\ this field is much weaker than that of the disk, and so may be easily missed if quantities are not plotted with care. 

Additionally, although neither simulation produces sufficient horizon-crossing magnetic flux to exceed the MAD transitional value of $\phi=50$ for a non-spinning BH, our initially-strong-field case \lowbeta\ comes close at $\phi\sim40$, much higher than is typically seen in SANE disks. In disks such as those considered here where the magnetic flux (and magnetic field strength though the magnetic plasma-$\beta$) appears saturated but the MRI simultaneously appears active, the question of SANE vs MAD relies heavily on the definition of a MAD state, for which there is still heated debate \citep[e.g.][]{BegelmanEA2022}.

Our fiducial high $\beta$ simulation \highbeta\ produced an overdensity at early times that persisted throughout the run. No comparable overdensity was observed in the high resolution counterpart \highbetasup. The impact of the overdensity on disk evolution is mixed: while there are clear changes in the disk scale height, there are no clearly observable differences in the magnetic field strength or structure according to the metrics used (magnetic plasma $\beta$ and magnetic flux $\Phi$). Nonetheless, this finding highlights the need for great care when choosing a simulation mesh, even at high resolutions.

Our results suggest that disk metrics like the magnetic plasma $\beta$ and magnetic flux $\Phi$ are not inherent to our disks but are instead influenced by the initial magnetic conditions--in our case, the initial strength of the toroidal field--supporting the conclusions made by \citet{WhiteEA2020}. Such a finding has implications for both simulated and real accretion disks. It has been previously hoped that, given enough time, a simulated accretion disk eventually forgets its initial conditions and settles to a common stable state. Previous work with NVF and ZNVF poloidal fields has suggested that this is not the case, and our results now also extend this to the purely toroidal case for RIAFs. It is also possible that different modes of material inflow (e.g. feeding via a binary companion vs the ambient medium) could lead to different final states in real accretion disks. Further work utilizing more realistic initial conditions and environments may help to quantify whether, and under what conditions, real RIAFs are sensitive to these changes.

\begin{acknowledgments}
PER thanks the Gates Cambridge Trust for research funding. This work was supported, in whole or in part, by the Bill \& Melinda Gates Foundation [OPP1144]. Under the grant conditions of the Foundation, a Creative Commons Attribution 4.0 Generic License has already been assigned to the Author Accepted Manuscript version that might arise from this submission. CSR thanks the STFC for support under the Consolidated Grant ST/S000623/1, as well as the European Research Council (ERC) for support under the European Union’s Horizon 2020 research and innovation programme (grant 834203).
\end{acknowledgments}

\vspace{5mm}
\software{Athena++ \citep{athena++},
          numpy \citep{numpy},
          pandas \citep{pandas}
          scipy \citep{scipy},
          VisIt \citep{VisIt}
          }

\bibliography{disk_dynamo}{}

\begin{thebibliography}{}
\expandafter\ifx\csname natexlab\endcsname\relax\def\natexlab#1{#1}\fi
\providecommand{\url}[1]{\href{#1}{#1}}
\providecommand{\dodoi}[1]{doi:~\href{http://doi.org/#1}{\nolinkurl{#1}}}
\providecommand{\doeprint}[1]{\href{http://ascl.net/#1}{\nolinkurl{http://ascl.net/#1}}}
\providecommand{\doarXiv}[1]{\href{https://arxiv.org/abs/#1}{\nolinkurl{https://arxiv.org/abs/#1}}}

\bibitem[{{Bai} \& {Stone}(2013)}]{BaiStone2013}
{Bai}, X.-N., \& {Stone}, J.~M. 2013, \apj, 767, 30, \dodoi{10.1088/0004-637X/767/1/30}

\bibitem[{{Balbus}(2003)}]{Balbus2003}
{Balbus}, S.~A. 2003, \araa, 41, 555, \dodoi{10.1146/annurev.astro.41.081401.155207}

\bibitem[{{Balbus} \& {Hawley}(1991)}]{BalbusHawley1991}
{Balbus}, S.~A., \& {Hawley}, J.~F. 1991, \apj, 376, 214, \dodoi{10.1086/170270}

\bibitem[{{Balbus} \& {Hawley}(1998)}]{BalbusHawley1998}
---. 1998, Reviews of Modern Physics, 70, 1, \dodoi{10.1103/RevModPhys.70.1}

\bibitem[{{Balbus} {et~al.}(1996){Balbus}, {Hawley}, \& {Stone}}]{BalbusEA1996}
{Balbus}, S.~A., {Hawley}, J.~F., \& {Stone}, J.~M. 1996, \apj, 467, 76, \dodoi{10.1086/177585}

\bibitem[{{Begelman} {et~al.}(2022){Begelman}, {Scepi}, \& {Dexter}}]{BegelmanEA2022}
{Begelman}, M.~C., {Scepi}, N., \& {Dexter}, J. 2022, \mnras, 511, 2040, \dodoi{10.1093/mnras/stab3790}

\bibitem[{{Blandford} \& {Payne}(1982)}]{BlandfordPayne1982}
{Blandford}, R.~D., \& {Payne}, D.~G. 1982, \mnras, 199, 883, \dodoi{10.1093/mnras/199.4.883}

\bibitem[{{Blandford} \& {Znajek}(1977)}]{BlandfordZnajek1977}
{Blandford}, R.~D., \& {Znajek}, R.~L. 1977, \mnras, 179, 433, \dodoi{10.1093/mnras/179.3.433}

\bibitem[{Childs {et~al.}(2012)Childs, Brugger, Whitlock, Meredith, Ahern, Pugmire, Biagas, Miller, Harrison, Weber, Krishnan, Fogal, Sanderson, Garth, Bethel, Camp, R\"{u}bel, Durant, Favre, \& Navr\'{a}til}]{VisIt}
Childs, H., Brugger, E., Whitlock, B., {et~al.} 2012, in High Performance Visualization--Enabling Extreme-Scale Scientific Insight, 357--372, \dodoi{10.1201/b12985}

\bibitem[{{Davis} \& {Tchekhovskoy}(2020)}]{DavisTchekhovskoy2020}
{Davis}, S.~W., \& {Tchekhovskoy}, A. 2020, \araa, 58, 407, \dodoi{10.1146/annurev-astro-081817-051905}

\bibitem[{{Evans} \& {Hawley}(1988)}]{EvansHawley1988}
{Evans}, C.~R., \& {Hawley}, J.~F. 1988, \apj, 332, 659, \dodoi{10.1086/166684}

\bibitem[{{Event Horizon Telescope Collaboration} {et~al.}(2019){Event Horizon Telescope Collaboration}, {Akiyama}, {Alberdi}, {Alef}, {Asada}, {Azulay}, {Baczko}, {Ball}, {Balokovi{\'c}}, {Barrett}, {Bintley}, {Blackburn}, {Boland}, {Bouman}, {Bower}, {Bremer}, {Brinkerink}, {Brissenden}, {Britzen}, {Broderick}, {Broguiere}, {Bronzwaer}, {Byun}, {Carlstrom}, {Chael}, {Chan}, {Chatterjee}, {Chatterjee}, {Chen}, {Chen}, {Cho}, {Christian}, {Conway}, {Cordes}, {Crew}, {Cui}, {Davelaar}, {De Laurentis}, {Deane}, {Dempsey}, {Desvignes}, {Dexter}, {Doeleman}, {Eatough}, {Falcke}, {Fish}, {Fomalont}, {Fraga-Encinas}, {Friberg}, {Fromm}, {G{\'o}mez}, {Galison}, {Gammie}, {Garc{\'\i}a}, {Gentaz}, {Georgiev}, {Goddi}, {Gold}, {Gu}, {Gurwell}, {Hada}, {Hecht}, {Hesper}, {Ho}, {Ho}, {Honma}, {Huang}, {Huang}, {Hughes}, {Ikeda}, {Inoue}, {Issaoun}, {James}, {Jannuzi}, {Janssen}, {Jeter}, {Jiang}, {Johnson}, {Jorstad}, {Jung}, {Karami}, {Karuppusamy}, {Kawashima}, {Keating}, {Kettenis}, {Kim}, {Kim}, {Kim}, {Kino},
  {Koay}, {Koch}, {Koyama}, {Kramer}, {Kramer}, {Krichbaum}, {Kuo}, {Lauer}, {Lee}, {Li}, {Li}, {Lindqvist}, {Liu}, {Liuzzo}, {Lo}, {Lobanov}, {Loinard}, {Lonsdale}, {Lu}, {MacDonald}, {Mao}, {Markoff}, {Marrone}, {Marscher}, {Mart{\'\i}-Vidal}, {Matsushita}, {Matthews}, {Medeiros}, {Menten}, {Mizuno}, {Mizuno}, {Moran}, {Moriyama}, {Moscibrodzka}, {Mul{\ensuremath{\ddot{}}}ler}, {Nagai}, {Nagar}, {Nakamura}, {Narayan}, {Narayanan}, {Natarajan}, {Neri}, {Ni}, {Noutsos}, {Okino}, {Olivares}, {Oyama}, {{\"O}zel}, {Palumbo}, {Patel}, {Pen}, {Pesce}, {Pi{\'e}tu}, {Plambeck}, {PopStefanija}, {Porth}, {Prather}, {Preciado-L{\'o}pez}, {Psaltis}, {Pu}, {Ramakrishnan}, {Rao}, {Rawlings}, {Raymond}, {Rezzolla}, {Ripperda}, {Roelofs}, {Rogers}, {Ros}, {Rose}, {Roshanineshat}, {Rottmann}, {Roy}, {Ruszczyk}, {Ryan}, {Rygl}, {S{\'a}nchez}, {S{\'a}nchez-Arguelles}, {Sasada}, {Savolainen}, {Schloerb}, {Schuster}, {Shao}, {Shen}, {Small}, {Sohn}, {SooHoo}, {Tazaki}, {Tiede}, {Tilanus}, {Titus}, {Toma}, {Torne}, {Trent},
  {Trippe}, {Tsuda}, {van Bemmel}, {van Langevelde}, {van Rossum}, {Wagner}, {Wardle}, {Weintroub}, {Wex}, {Wharton}, {Wielgus}, {Wong}, {Wu}, {Young}, {Young}, {Younsi}, {Yuan}, {Yuan}, {Zensus}, {Zhao}, {Zhao}, {Zhu}, {Anczarski}, {Baganoff}, {Eckart}, {Farah}, {Haggard}, {Meyer-Zhao}, {Michalik}, {Nadolski}, {Neilsen}, {Nishioka}, {Nowak}, {Pradel}, {Primiani}, {Souccar}, {Vertatschitsch}, {Yamaguchi}, \& {Zhang}}]{EHTCollab2019}
{Event Horizon Telescope Collaboration}, {Akiyama}, K., {Alberdi}, A., {et~al.} 2019, \apjl, 875, L5, \dodoi{10.3847/2041-8213/ab0f43}

\bibitem[{{Falanga} {et~al.}(2007){Falanga}, {Melia}, {Tagger}, {Goldwurm}, \& {B{\'e}langer}}]{FalangaEA2007}
{Falanga}, M., {Melia}, F., {Tagger}, M., {Goldwurm}, A., \& {B{\'e}langer}, G. 2007, \apjl, 662, L15, \dodoi{10.1086/519278}

\bibitem[{{Fishbone} \& {Moncrief}(1976)}]{FishboneMoncrief1976}
{Fishbone}, L.~G., \& {Moncrief}, V. 1976, \apj, 207, 962, \dodoi{10.1086/154565}

\bibitem[{{Flock} {et~al.}(2010){Flock}, {Dzyurkevich}, {Klahr}, \& {Mignone}}]{FlockEA2010}
{Flock}, M., {Dzyurkevich}, N., {Klahr}, H., \& {Mignone}, A. 2010, \aap, 516, A26, \dodoi{10.1051/0004-6361/200912443}

\bibitem[{{Fragile} \& {S{\k{a}}dowski}(2017)}]{FragileSadowski2017}
{Fragile}, P.~C., \& {S{\k{a}}dowski}, A. 2017, \mnras, 467, 1838, \dodoi{10.1093/mnras/stx274}

\bibitem[{{Giannios} \& {Uzdensky}(2019)}]{GianniosUzdensky2019}
{Giannios}, D., \& {Uzdensky}, D.~A. 2019, \mnras, 484, 1378, \dodoi{10.1093/mnras/stz082}

\bibitem[{{Goodman} \& {Xu}(1994)}]{GoodmanXu1994}
{Goodman}, J., \& {Xu}, G. 1994, \apj, 432, 213, \dodoi{10.1086/174562}

\bibitem[{{Gressel} \& {Pessah}(2015)}]{GresselPessah2015}
{Gressel}, O., \& {Pessah}, M.~E. 2015, \apj, 810, 59, \dodoi{10.1088/0004-637X/810/1/59}

\bibitem[{{Guan} {et~al.}(2009){Guan}, {Gammie}, {Simon}, \& {Johnson}}]{GuanEA2009}
{Guan}, X., {Gammie}, C.~F., {Simon}, J.~B., \& {Johnson}, B.~M. 2009, \apj, 694, 1010, \dodoi{10.1088/0004-637X/694/2/1010}

\bibitem[{Harris {et~al.}(2020)Harris, Millman, van~der Walt, Gommers, Virtanen, Cournapeau, Wieser, Taylor, Berg, Smith, Kern, Picus, Hoyer, van Kerkwijk, Brett, Haldane, del R{\'{i}}o, Wiebe, Peterson, G{\'{e}}rard-Marchant, Sheppard, Reddy, Weckesser, Abbasi, Gohlke, \& Oliphant}]{numpy}
Harris, C.~R., Millman, K.~J., van~der Walt, S.~J., {et~al.} 2020, Nature, 585, 357, \dodoi{10.1038/s41586-020-2649-2}

\bibitem[{{Hawley} {et~al.}(1995){Hawley}, {Gammie}, \& {Balbus}}]{HawleyEA1995}
{Hawley}, J.~F., {Gammie}, C.~F., \& {Balbus}, S.~A. 1995, \apj, 440, 742, \dodoi{10.1086/175311}

\bibitem[{{Hawley} {et~al.}(1996){Hawley}, {Gammie}, \& {Balbus}}]{HawleyEA1996}
---. 1996, \apj, 464, 690, \dodoi{10.1086/177356}

\bibitem[{{Hawley} {et~al.}(2011){Hawley}, {Guan}, \& {Krolik}}]{HawleyEA2011}
{Hawley}, J.~F., {Guan}, X., \& {Krolik}, J.~H. 2011, \apj, 738, 84, \dodoi{10.1088/0004-637X/738/1/84}

\bibitem[{{Hawley} {et~al.}(2013){Hawley}, {Richers}, {Guan}, \& {Krolik}}]{HawleyEA2013}
{Hawley}, J.~F., {Richers}, S.~A., {Guan}, X., \& {Krolik}, J.~H. 2013, \apj, 772, 102, \dodoi{10.1088/0004-637X/772/2/102}

\bibitem[{{Hogg} \& {Reynolds}(2016)}]{HoggReynolds2016}
{Hogg}, J.~D., \& {Reynolds}, C.~S. 2016, \apj, 826, 40, \dodoi{10.3847/0004-637X/826/1/40}

\bibitem[{{Hogg} \& {Reynolds}(2018{\natexlab{a}})}]{HoggReynolds2018A}
---. 2018{\natexlab{a}}, \apj, 854, 6, \dodoi{10.3847/1538-4357/aaa6c6}

\bibitem[{{Hogg} \& {Reynolds}(2018{\natexlab{b}})}]{HoggReynolds2018B}
---. 2018{\natexlab{b}}, \apj, 861, 24, \dodoi{10.3847/1538-4357/aac439}

\bibitem[{{Jiang} {et~al.}(2019){Jiang}, {Stone}, \& {Davis}}]{JiangEA2019}
{Jiang}, Y.-F., {Stone}, J.~M., \& {Davis}, S.~W. 2019, \apj, 880, 67, \dodoi{10.3847/1538-4357/ab29ff}

\bibitem[{{King}(2003)}]{King2003}
{King}, A. 2003, \apjl, 596, L27, \dodoi{10.1086/379143}

\bibitem[{{Liska} {et~al.}(2020){Liska}, {Tchekhovskoy}, \& {Quataert}}]{LiskaEA2020}
{Liska}, M., {Tchekhovskoy}, A., \& {Quataert}, E. 2020, \mnras, 494, 3656, \dodoi{10.1093/mnras/staa955}

\bibitem[{{Lovelace} {et~al.}(1999){Lovelace}, {Li}, {Colgate}, \& {Nelson}}]{LovelaceEA1999}
{Lovelace}, R.~V.~E., {Li}, H., {Colgate}, S.~A., \& {Nelson}, A.~F. 1999, \apj, 513, 805, \dodoi{10.1086/306900}

\bibitem[{{Machida} {et~al.}(2001){Machida}, {Matsumoto}, \& {Mineshige}}]{MachidaEA2001}
{Machida}, M., {Matsumoto}, R., \& {Mineshige}, S. 2001, \pasj, 53, L1, \dodoi{10.1093/pasj/53.1.L1}

\bibitem[{{Mignon-Risse} {et~al.}(2021){Mignon-Risse}, {Aimar}, {Varniere}, {Casse}, \& {Vincent}}]{Mignon-RisseEA2021}
{Mignon-Risse}, R., {Aimar}, N., {Varniere}, P., {Casse}, F., \& {Vincent}, F. 2021, in SF2A-2021: Proceedings of the Annual meeting of the French Society of Astronomy and Astrophysics. Eds.: A. Siebert, ed. A.~{Siebert}, K.~{Bailli{\'e}}, E.~{Lagadec}, N.~{Lagarde}, J.~{Malzac}, J.~B. {Marquette}, M.~{N'Diaye}, J.~{Richard}, \& O.~{Venot}, 113--116

\bibitem[{{Mishra} {et~al.}(2020){Mishra}, {Begelman}, {Armitage}, \& {Simon}}]{MishraEA2020}
{Mishra}, B., {Begelman}, M.~C., {Armitage}, P.~J., \& {Simon}, J.~B. 2020, \mnras, 492, 1855, \dodoi{10.1093/mnras/stz3572}

\bibitem[{{Mishra} {et~al.}(2022){Mishra}, {Fragile}, {Anderson}, {Blankenship}, {Li}, \& {Nalewajko}}]{MishraEA2022}
{Mishra}, B., {Fragile}, P.~C., {Anderson}, J., {et~al.} 2022, \apj, 939, 31, \dodoi{10.3847/1538-4357/ac938b}

\bibitem[{{Narayan} {et~al.}(2003){Narayan}, {Igumenshchev}, \& {Abramowicz}}]{NarayanEA2003}
{Narayan}, R., {Igumenshchev}, I.~V., \& {Abramowicz}, M.~A. 2003, \pasj, 55, L69, \dodoi{10.1093/pasj/55.6.L69}

\bibitem[{{Narayan} {et~al.}(2012){Narayan}, {S{\k{a}}dowski}, {Penna}, \& {Kulkarni}}]{NarayanEA2012}
{Narayan}, R., {S{\k{a}}dowski}, A., {Penna}, R.~F., \& {Kulkarni}, A.~K. 2012, \mnras, 426, 3241, \dodoi{10.1111/j.1365-2966.2012.22002.x}

\bibitem[{{Noble} {et~al.}(2010){Noble}, {Krolik}, \& {Hawley}}]{NobleEA2010}
{Noble}, S.~C., {Krolik}, J.~H., \& {Hawley}, J.~F. 2010, \apj, 711, 959, \dodoi{10.1088/0004-637X/711/2/959}

\bibitem[{{Papaloizou} \& {Pringle}(1984)}]{PapaloizouPringle1984}
{Papaloizou}, J.~C.~B., \& {Pringle}, J.~E. 1984, \mnras, 208, 721, \dodoi{10.1093/mnras/208.4.721}

\bibitem[{{Penna} {et~al.}(2013){Penna}, {S{\k{a}}dowski}, {Kulkarni}, \& {Narayan}}]{PennaEA2013}
{Penna}, R.~F., {S{\k{a}}dowski}, A., {Kulkarni}, A.~K., \& {Narayan}, R. 2013, \mnras, 428, 2255, \dodoi{10.1093/mnras/sts185}

\bibitem[{{Pessah}(2010)}]{Pessah2010}
{Pessah}, M.~E. 2010, \apj, 716, 1012, \dodoi{10.1088/0004-637X/716/2/1012}

\bibitem[{{Rincon} {et~al.}(2007){Rincon}, {Ogilvie}, \& {Proctor}}]{Rincon2007}
{Rincon}, F., {Ogilvie}, G.~I., \& {Proctor}, M.~R.~E. 2007, \prl, 98, 254502, \dodoi{10.1103/PhysRevLett.98.254502}

\bibitem[{{Salvesen} {et~al.}(2016){Salvesen}, {Simon}, {Armitage}, \& {Begelman}}]{SalvesenEA2016}
{Salvesen}, G., {Simon}, J.~B., {Armitage}, P.~J., \& {Begelman}, M.~C. 2016, \mnras, 457, 857, \dodoi{10.1093/mnras/stw029}

\bibitem[{{Shakura} \& {Sunyaev}(1973)}]{ShakuraSunyaev1973}
{Shakura}, N.~I., \& {Sunyaev}, R.~A. 1973, \aap, 24, 337

\bibitem[{{Sorathia} {et~al.}(2010){Sorathia}, {Reynolds}, \& {Armitage}}]{SorathiaEA2010}
{Sorathia}, K.~A., {Reynolds}, C.~S., \& {Armitage}, P.~J. 2010, \apj, 712, 1241, \dodoi{10.1088/0004-637X/712/2/1241}

\bibitem[{{Sorathia} {et~al.}(2012){Sorathia}, {Reynolds}, {Stone}, \& {Beckwith}}]{SorathiaEA2012}
{Sorathia}, K.~A., {Reynolds}, C.~S., {Stone}, J.~M., \& {Beckwith}, K. 2012, \apj, 749, 189, \dodoi{10.1088/0004-637X/749/2/189}

\bibitem[{{Stone} {et~al.}(2008){Stone}, {Gardiner}, {Teuben}, {Hawley}, \& {Simon}}]{athena}
{Stone}, J.~M., {Gardiner}, T.~A., {Teuben}, P., {Hawley}, J.~F., \& {Simon}, J.~B. 2008, \apjs, 178, 137, \dodoi{10.1086/588755}

\bibitem[{{Stone} {et~al.}(2020){Stone}, {Tomida}, {White}, \& {Felker}}]{athena++}
{Stone}, J.~M., {Tomida}, K., {White}, C.~J., \& {Felker}, K.~G. 2020, \apjs, 249, 4, \dodoi{10.3847/1538-4365/ab929b}

\bibitem[{{Tchekhovskoy} {et~al.}(2012){Tchekhovskoy}, {McKinney}, \& {Narayan}}]{TchekhovskoyEA2012}
{Tchekhovskoy}, A., {McKinney}, J.~C., \& {Narayan}, R. 2012, in Journal of Physics Conference Series, Vol. 372, Journal of Physics Conference Series, 012040, \dodoi{10.1088/1742-6596/372/1/012040}

\bibitem[{{Tchekhovskoy} {et~al.}(2011){Tchekhovskoy}, {Narayan}, \& {McKinney}}]{TchekhovskoyEA2011}
{Tchekhovskoy}, A., {Narayan}, R., \& {McKinney}, J.~C. 2011, \mnras, 418, L79, \dodoi{10.1111/j.1745-3933.2011.01147.x}

\bibitem[{Virtanen {et~al.}(2020)Virtanen, Gommers, Oliphant, Haberland, Reddy, Cournapeau, Burovski, Peterson, Weckesser, Bright, {van der Walt}, Brett, Wilson, Millman, Mayorov, Nelson, Jones, Kern, Larson, Carey, Polat, Feng, Moore, {VanderPlas}, Laxalde, Perktold, Cimrman, Henriksen, Quintero, Harris, Archibald, Ribeiro, Pedregosa, {van Mulbregt}, \& {SciPy 1.0 Contributors}}]{scipy}
Virtanen, P., Gommers, R., Oliphant, T.~E., {et~al.} 2020, Nature Methods, 17, 261, \dodoi{10.1038/s41592-019-0686-2}

\bibitem[{{W}es {M}c{K}inney(2010)}]{pandas}
{W}es {M}c{K}inney. 2010, in {P}roceedings of the 9th {P}ython in {S}cience {C}onference, ed. {S}t\'efan van~der {W}alt \& {J}arrod {M}illman, 56 -- 61, \dodoi{10.25080/Majora-92bf1922-00a}

\bibitem[{{White} {et~al.}(2020){White}, {Quataert}, \& {Gammie}}]{WhiteEA2020}
{White}, C.~J., {Quataert}, E., \& {Gammie}, C.~F. 2020, \apj, 891, 63, \dodoi{10.3847/1538-4357/ab718e}

\end{thebibliography}
\bibliographystyle{aasjournal}

\end{document}